\documentclass{article}

\usepackage{microtype}
\usepackage{graphicx}
\usepackage{subfigure}
\usepackage{booktabs} 
\usepackage{amsmath}
\usepackage{amssymb}
\usepackage{bm}
\usepackage{algorithmic}
\usepackage{booktabs}
\usepackage{natbib}
\usepackage{mathtools}
\usepackage{upgreek}
\usepackage{makeidx}
\usepackage{tikz}
\usepackage{standalone}

\usepackage{tabularx}

\usepackage{ml-defs}

\usepackage{hyperref}

\usepackage[accepted]{icml2020}

\icmltitlerunning{Large-scale ligand-based virtual screening 
for SARS-CoV-2 inhibitors using deep neural 
networks}

\begin{document}

\twocolumn[
\icmltitle{Large-scale ligand-based virtual screening for SARS-CoV-2 inhibitors using deep neural networks}



\icmlsetsymbol{equal}{*}

\begin{icmlauthorlist}

\icmlauthor{Markus Hofmarcher}{lit,iml,equal}
\icmlauthor{Andreas Mayr}{lit,iml,equal}
\icmlauthor{Elisabeth Rumetshofer}{lit,iml,equal}
\icmlauthor{Peter Ruch}{lit,iml,equal}
\icmlauthor{Philipp Renz}{lit,iml,equal}
\icmlauthor{Johannes Schimunek}{lit,iml,equal}
\icmlauthor{Philipp Seidl}{lit,iml,equal}
\icmlauthor{Andreu Vall}{lit,iml,equal}
\icmlauthor{Michael Widrich}{lit,iml,equal}
\icmlauthor{Sepp Hochreiter}{lit,iml,equal}
\icmlauthor{G{\"u}nter Klambauer}{lit,iml,equal}

\end{icmlauthorlist}

\icmlaffiliation{lit}{ELLIS Unit at the LIT AI Lab, Johannes Kepler University Linz, Austria}
\icmlaffiliation{iml}{Institute for Machine Learning, Johannes Kepler University Linz, Austria}

\icmlcorrespondingauthor{G{\"u}nter Klambauer}{klambauer@ml.jku.at}

\icmlkeywords{Artificial intelligence, neural networks, deep learning,
QSAR, virtual screening, SARS-CoV, Corona, SARS-CoV-2,
Cov inhibitors}
\vskip 0.3in
]



\printAffiliationsAndNotice{\icmlEqualContribution} 

\begin{abstract}
Due to the current severe acute respiratory syndrome coronavirus 2 (SARS-CoV-2) 
pandemic, there is an urgent need for novel therapies and drugs. We conducted 
a large-scale virtual screening for small molecules that are potential CoV-2 
inhibitors. To this end, we utilized ``ChemAI,'' a deep neural network trained 
on more than 220M 
data points across 3.6M molecules from three public drug-discovery databases. 
With ChemAI, we screened and ranked one billion molecules from the ZINC database for favourable 
effects against CoV-2. We then reduced the result to the 30,000 top-ranked compounds, 
which are readily accessible and purchasable via the ZINC database. 
Additionally, we screened the DrugBank using ChemAI to allow for drug repurposing, which
would be a fast way towards a therapy. We provide these 
top-ranked compounds of ZINC and DrugBank as a library for 
further screening with bioassays 
at \url{https://github.com/ml-jku/sars-cov-inhibitors-chemai}.
\end{abstract}

\paragraph{Introduction.}
Due to the current world-wide crisis of SARS-CoV-2 virus infections, 
there is a strong need for new therapies.
While many efforts are focused on
repurposing existing drugs \citep{zhou2020network,wang2020virtual,ton2020rapid},
we suggest to test new molecules with potentially higher efficacy.
Therefore, we performed a large-scale ligand-based virtual 
screening run, which resulted in 30,000 potential SARS-CoV-2 inhibitors
with favorable properties. The screening method 
is at the core a deep neural network for 
drug discovery \citep{hochreiter2018machine,klambauer2019machine}. We 
actively reach out to the scientific
community to test these molecules and consider them as a 
custom-designed chemical library.

Most current virtual screens are structure-based and use docking 
methods \citep{chen2020prediction,huang2020virtual,haider2020silico,wang2020virtual,fischer2020inhibitors,chen2020prediction,ton2020rapid,senathilake2020virtual,ruan2020potential,jin2020structure,zhang2020discovery,gorgulla2020open} while only one 
screen is ligand-based and uses a similarity-based approach \citep{zhu2020d3similarity}.
The largest docking studies screen databases with sizes ranging 
from roughly 700 million \citep{fischer2020inhibitors} to 1.3 billion \citep{ton2020rapid} 
molecules. Also our study operates on databases of this size, concretely 
we perform a ligand-based virtual screening of a collection of one 
billion molecules from the ZINC database.

\begin{table*}[t]
    \centering
    \begin{tabular}{rlrrll}
        \toprule
         Assay ID &	Source		& \#inact	& \#act	& Description \\
        \midrule
        1706 	& PubChem		& 193637	& 269 & QFRET-based assay for SARS-CoV 3C-like Protease\\
        1879	& PubChem  	& 167	& 86	& QFRET-based assay for SARS-CoV 3C-like Protease (confirmation) \\
        485353	& PubChem		& 215030	& 390	& Yeast-based Assay for SARS-CoV PLP \\
        652038	& PubChem 	& 493	& 135	& Yeast-based Assay for SARS-CoV PLP (validation)  \\
        \bottomrule
    \end{tabular}
    \caption{Overview of the main biological effects considered for ranking the molecules of the virtual 
    screen. ``\#inact'' and ``\#actAll'' report the number of actives and inactives in the training set. All assays are based on inhibition of proteins of SARS-CoV-1.}
    \label{tab:assays}
\end{table*}

\paragraph{Deep ligand-based virtual screening.}
``ChemAI'' is a deep neural network trained to simultaneously predict a 
large number of biological effects~\citep{mayr2018large,preuer2019interpretable}. 
In more detail, the network is of the type 
SmilesLSTM \citep{mayr2018large,hochreiter1997long} and trained on a data set comprised of 
ChEMBL \citep{gaulton2017chembl}, ZINC \citep{sterling2015zinc} and 
PubChem \citep{kim2016pubchem}, and which is similar to the data set used 
by \citet{preuer2018frechet}. ChemAI predicts 6,269 biological outcomes, 
such as binding to targets, inhibitory or toxic effects. The network
was trained in a multi-task setting, in which data from other bioassays
was used to enhance the predictive power for SARS-CoV inhibitory effects.
Each modelled biological effect is represented by an output neuron 
of the neural network. We utilized a small set of output neurons 
associated with SARS-CoV inhibition and a set of output neurons
associated with toxic effects to rank compounds. 

We screen the ZINC database because it contains a large set of diverse 
molecules and additionally provides links to vendors from which to 
purchase and physically obtain those molecules. We 
downloaded 898,196,375 molecules from ZINC and converted them to
canonical SMILES \citep{weininger1988smiles} using RDKit \citep{2006rdkit}. 
We then performed 
inference with ChemAI to obtain predictions for 
each of those roughly one billion molecules.

\paragraph{Selecting bioassays for multiple targets of SARS-CoV.} 
The SARS-CoV-2 has two main proteases that 
are critical for its replication, namely the 3CLpro (3C-like protease) and PLpro (Papain 
Like Protease), encoded in an open reading frame 
\citep{macchiagodena2020inhibition}. A compound that inhibits 
both proteases could be promising drug candidates \citep{ledford2009one,collison2019two}. 
The virus proteases are also strikingly similar to those in SARS-CoV-1 
\citep{macchiagodena2020inhibition}, which is also an 
implicit assumption by docking-based approaches.
We therefore select two groups of assays, one of which measures
the inhibition of 3CLpro and the other
the inhibition of PLpro (see Table~\ref{tab:assays}). 
For each of those four assays, ChemAI possesses an 
output unit, which models the ability of small molecules
to exhibit the effect measured by the assay. Thus, 
using the predictions yielded by ChemAI, it is possible to rank
compounds by their predicted ability to inhibit the
two main proteases of SARS-CoV-1, which can be a proxy 
for the inhibitory potential for SARS-CoV-2.

\begin{table*}
\centering
\begin{tabular}{llrrrr}
\toprule
          ZINC ID &                                             Canonical SMILES &  dist &  score &  tox & ct\\
\midrule
 ZINC000254565785 &                CNC(=S)NN=Cc1c2ccccc2c(Cl)c2ccccc12 &             0.5455 &          0.8244 &           8 & 0.06\\
 ZINC000726422572 &         C=C(Cl)COc1ccc(C(C)=NNC(=S)NCCc2ccccn2)cc1 &             0.5333 &          0.8232 &           7 & 0.05\\
 ZINC000916265995 &                     CNC(=S)NN=Cc1cc2cccc(C)c2nc1Cl &             0.6111 &          0.8230 &           5 & 0.08\\
 ZINC000916356873 &       N\#CCCn1cc(C=NNC(=S)NCCc2ccc(Cl)cc2)c2ccccc21 &             0.6377 &          0.8221 &          17 & 0.07\\
 ZINC000806591744 &              O=c1c(Br)nn(Cc2cnc3ccccc3c2)c2ccccc12 &             0.7258 &          0.8215 &          11 & 0.16\\
 ZINC000178971373 &               O=c1c(Br)nn(Cc2nc3ccccc3s2)c2ccccc12 &             0.7288 &          0.8211 &           8 & 0.05\\
 ZINC000000155607 &                  CSC(=S)N/N=C/c1ccc2cc3ccccc3cc2c1 &             0.3902 &          0.8204 &           4 & 0.05\\
 ZINC000016317677 &    C=CCNC(=S)NNC(=O)Cn1c(COc2ccc(Cl)cc2)nc2ccccc21 &             0.7000 &          0.8197 &           4 & 0.07\\
 ZINC000193073749 &                 O=C(Cn1cccc(Br)c1=O)c1ccc2ccccc2c1 &             0.6667 &          0.8197 &           1 & 0.13\\
 ZINC000769846795 &              O=c1c(Br)nn(Cc2ccc3ncccc3c2)c2ccccc12 &             0.6949 &          0.8195 &           9 & 0.14\\
 ZINC000755523869 &              CN(N=Cc1nc2ccccn2c1Br)C(=S)NCc1ccccc1 &             0.6452 &          0.8194 &           4 & 0.05\\
 ZINC000763345954 &                   C=CCNC(=S)NN=Cc1nc2ccc(Cl)cc2n1C &             0.6508 &          0.8194 &           5 & 0.07\\
 ZINC000001448699 &          CSC(=S)N/N=C/c1nc(-c2ccc(Cl)cc2)n2ccccc12 &             0.6866 &          0.8192 &           4 & 0.13\\
 ZINC000016940508 &    C/C(=N\textbackslash NC(=S)NNC(=S)N(C)c1ccccc1)c1nccc2ccccc12 &             0.6721 &          0.8191 &          11 & 0.06\\
 ZINC000005486767 &    C/C(=N/NC(=S)NNC(=S)N(C)c1ccccc1)c1nccc2ccccc12 &             0.6721 &          0.8191 &          11 & 0.06\\
 ZINC000005527649 &                        CSC(=S)N/N=C/c1ccc2ccccc2n1 &             0.6327 &          0.8187 &           6 & 0.05\\
 ZINC000755497029 &                   C=CCNC(=S)NN=Cc1nc2cc(Cl)ccc2n1C &             0.6719 &          0.8186 &           5 & 0.06\\
 ZINC000746495682 &        FC(F)(F)CNC(=S)NN=Cc1cn(Cc2ccccc2)c2ccccc12 &             0.5690 &          0.8186 &          15 & 0.07\\
 ZINC000005719506 &                 CN(/N=C/c1ccc(Cl)cc1)C(=S)c1ccccc1 &             0.6818 &          0.8178 &           4 & 0.05\\
 ZINC000002149503 &  S=C(NCc1ccccc1)N/N=C/c1cn(CCOc2ccc(Br)cc2)c2ccccc12 &             0.5625 &          0.8175 &          13 & 0.21\\
\bottomrule
\end{tabular}

\caption{Top-ranked molecules by ChemAI. All compounds have a high 
activity predicted on all four assays (column ``score'') and are relatively 
distant (column ``dist'') to current known inhibitors. The distance measure
is the Jaccard distance based on binary ECFP4 fingerprints and resides
in the interval $[0,1]$. Some of the
presented molecules might exhibit a number of toxic effects (column ``tox'').
Here the number of models indicating a toxic effect is reported, where 
the total number of toxicity models was 75. 
We also report the estimated probability to exhibit clinical toxicity (column ``ct'').}
\label{tab:some_label}

\end{table*}


\begin{figure*}
    \centering
    \begin{tabular}{ccccc}
    
\includegraphics[width=32mm]{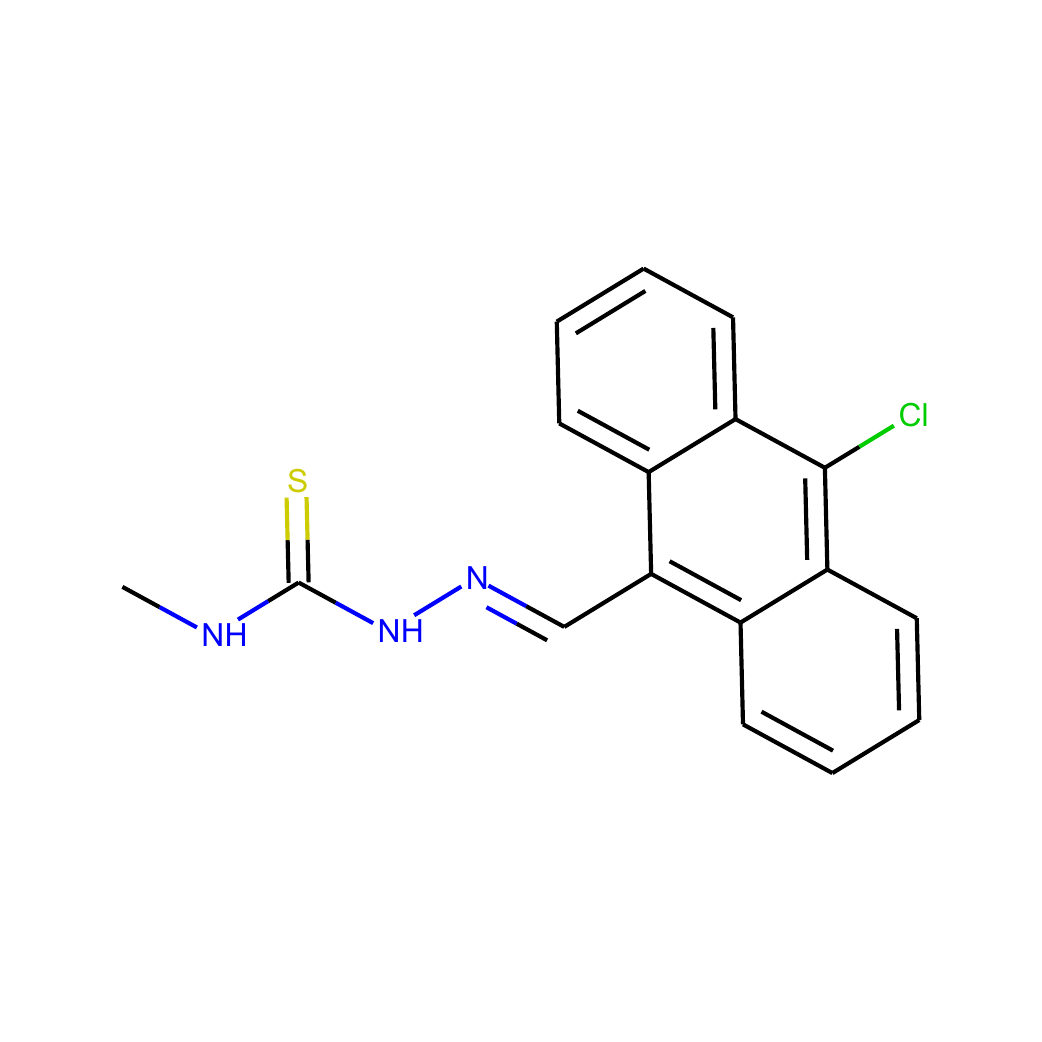} & \includegraphics[width=32mm]{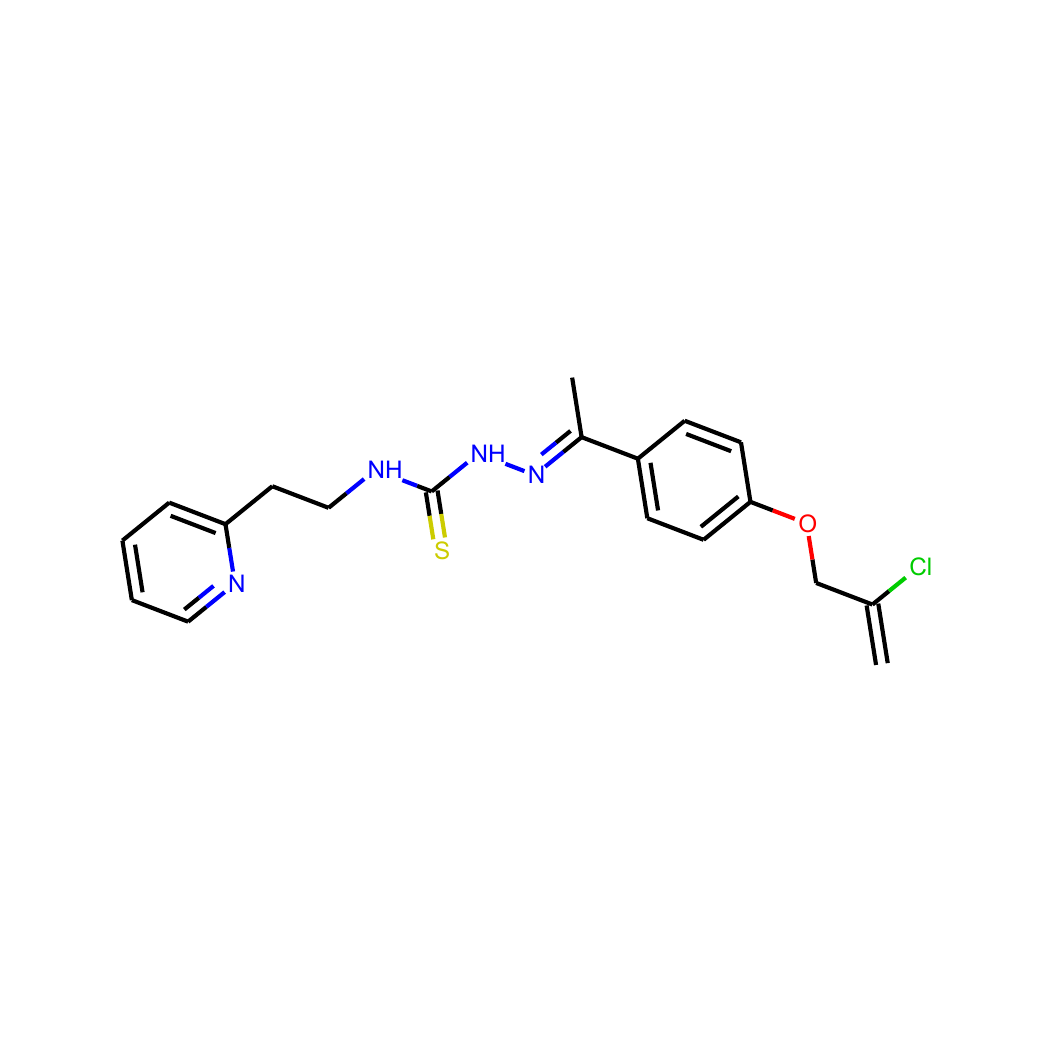} & \includegraphics[width=32mm]{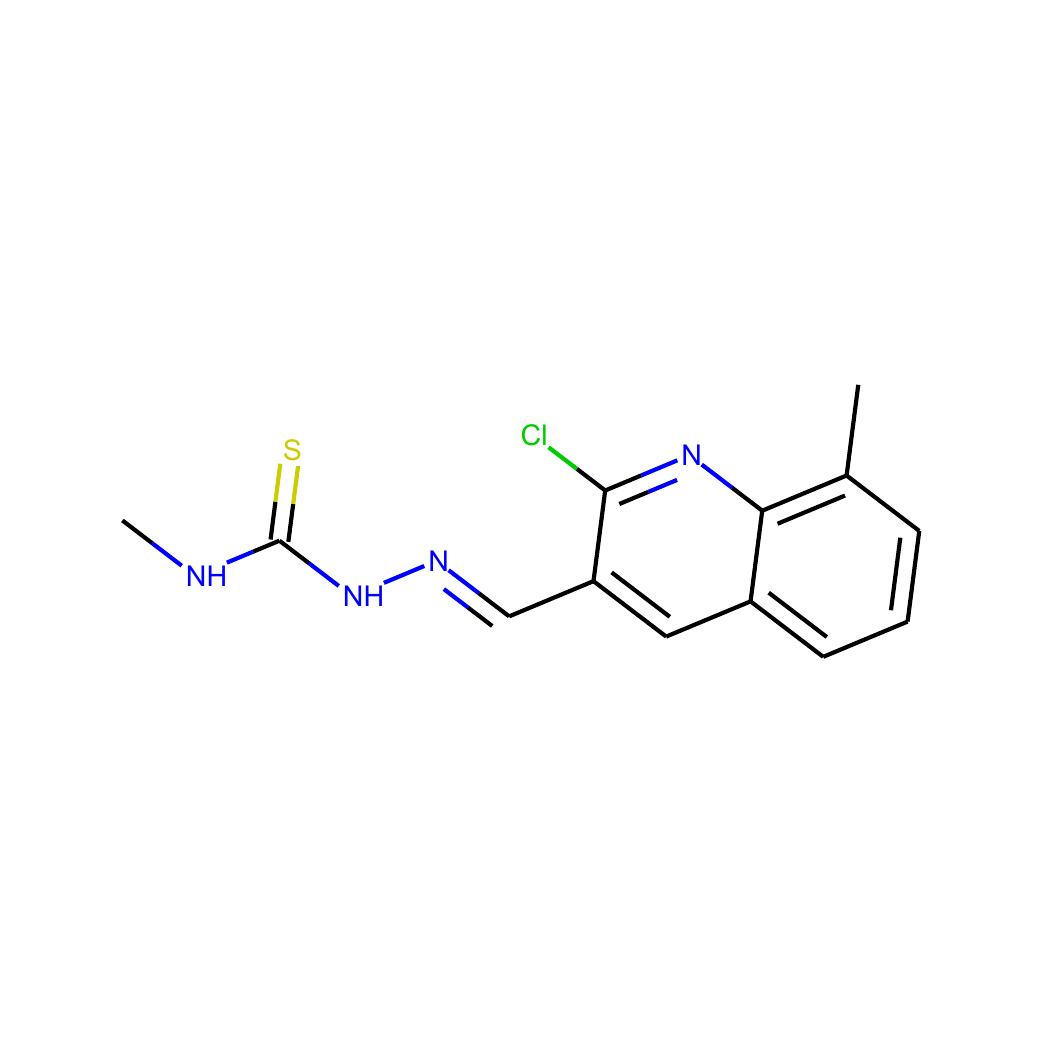} & \includegraphics[width=32mm]{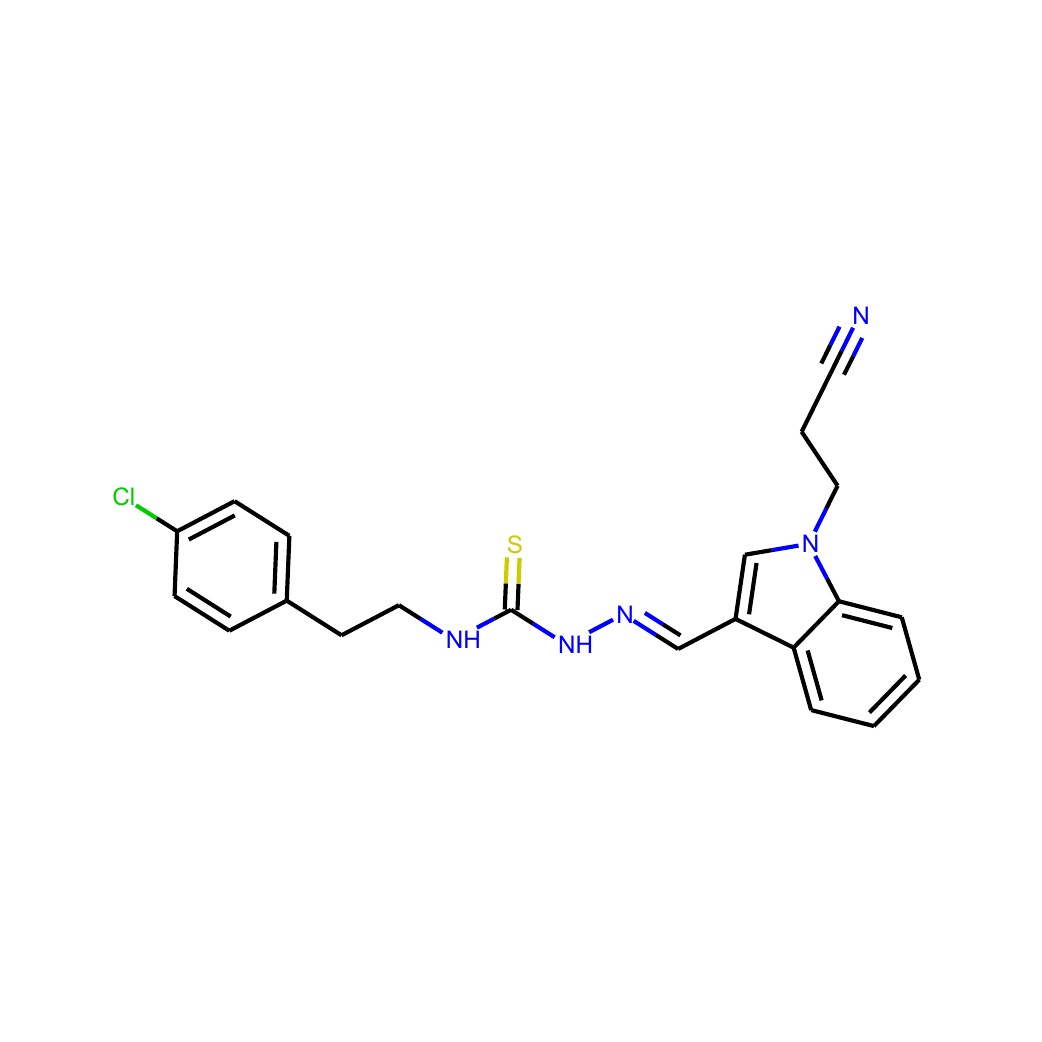} & \\
\href{https://zinc.docking.org/substances/ZINC000254565785/}{ZINC000254565785} & \href{https://zinc.docking.org/substances/ZINC000726422572/}{ZINC000726422572} & \href{https://zinc.docking.org/substances/ZINC000916265995/}{ZINC000916265995} & \href{https://zinc.docking.org/substances/ZINC000916356873/}{ZINC000916356873} \\
\includegraphics[width=32mm]{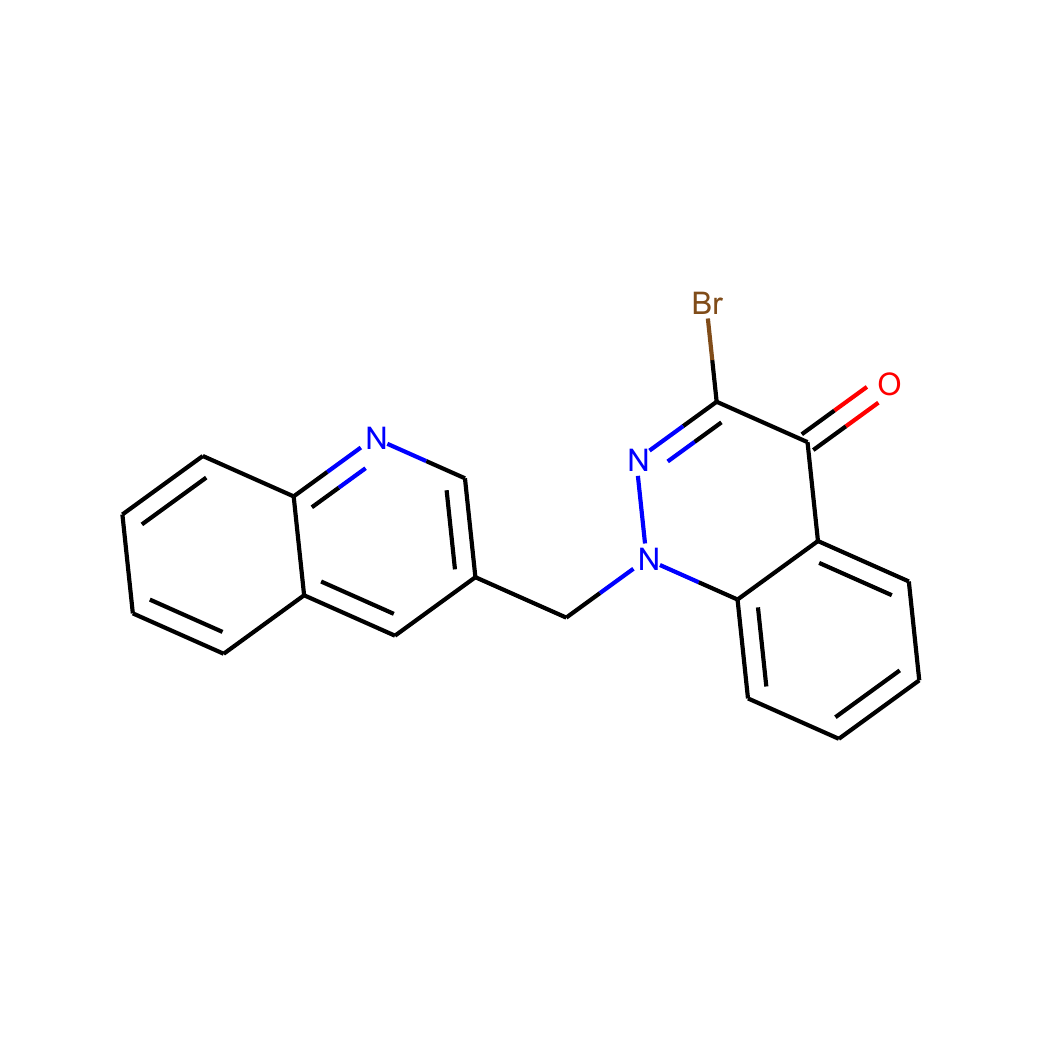} & \includegraphics[width=32mm]{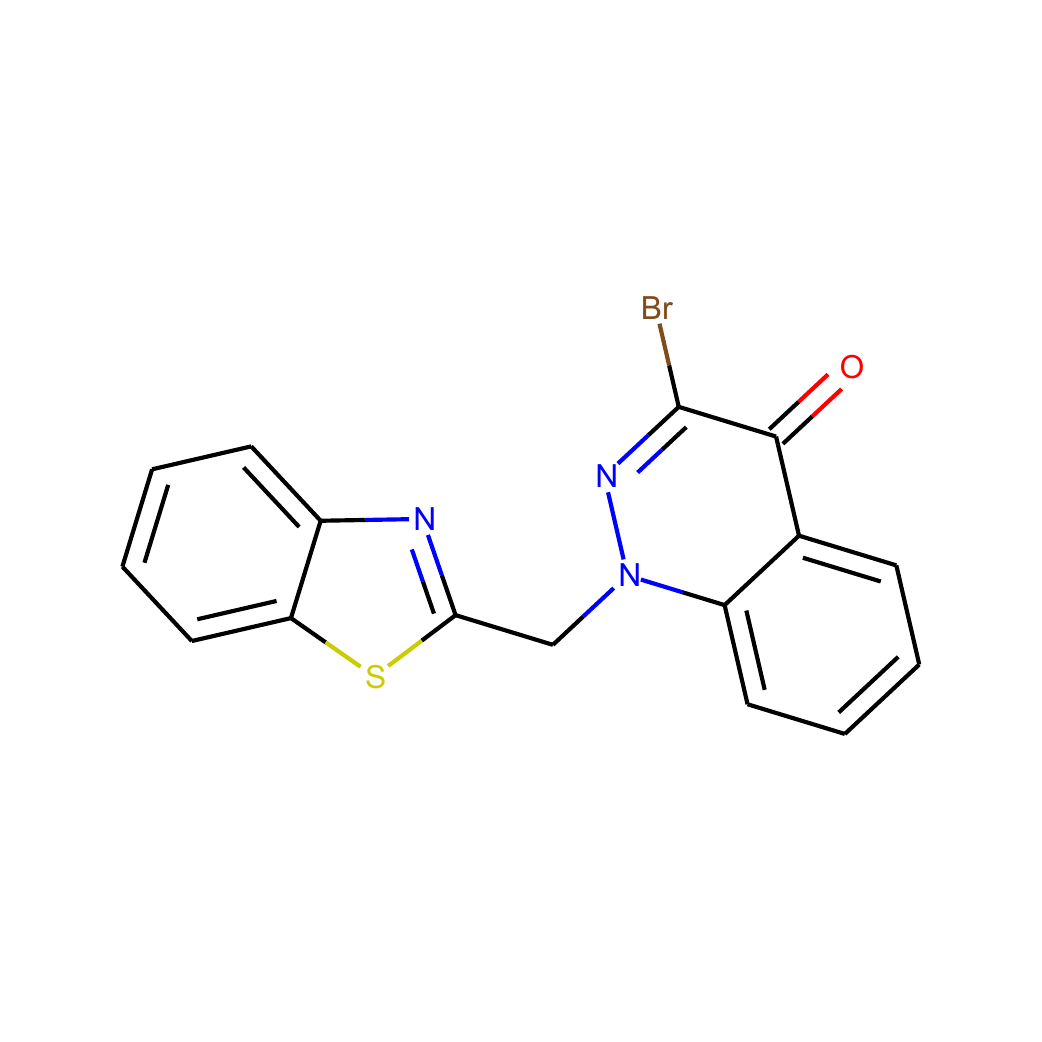} & \includegraphics[width=32mm]{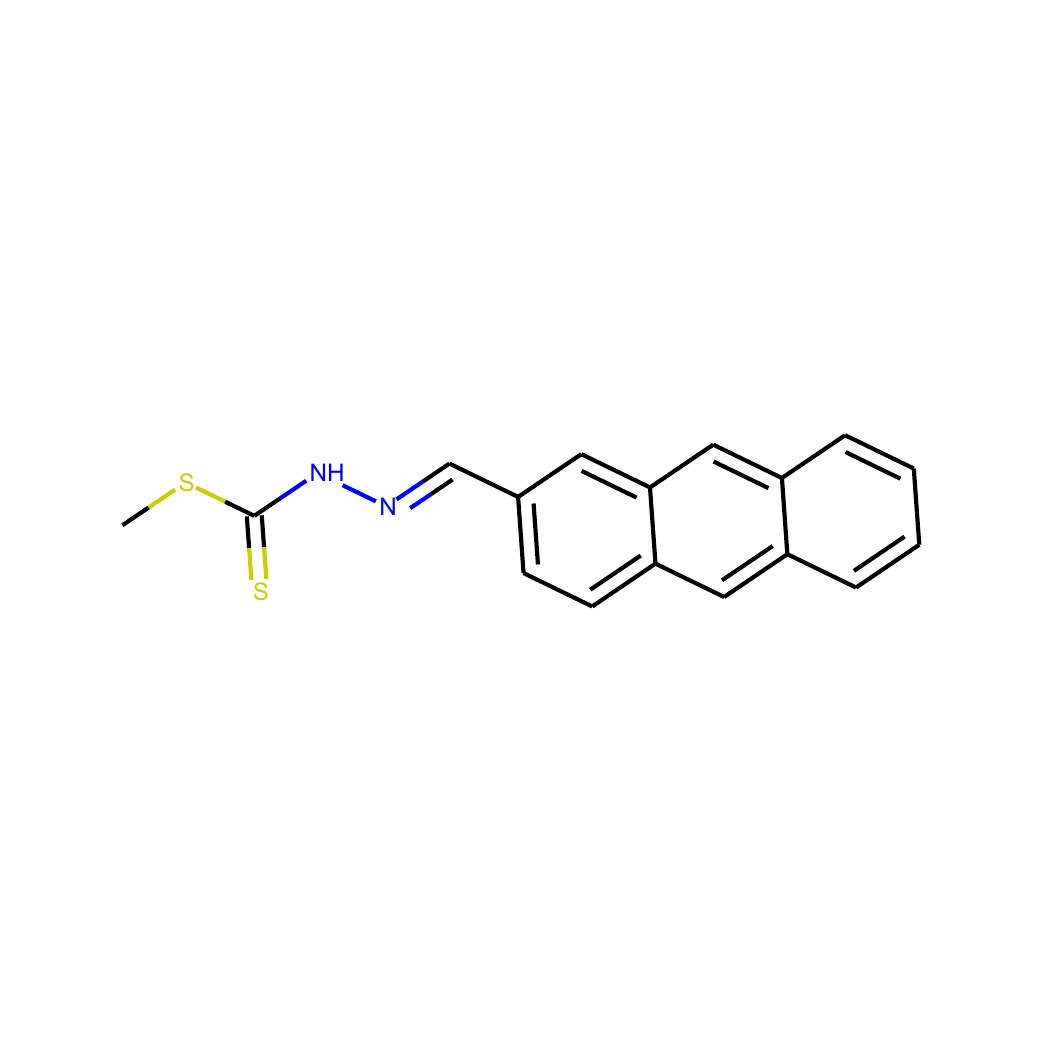} & \includegraphics[width=32mm]{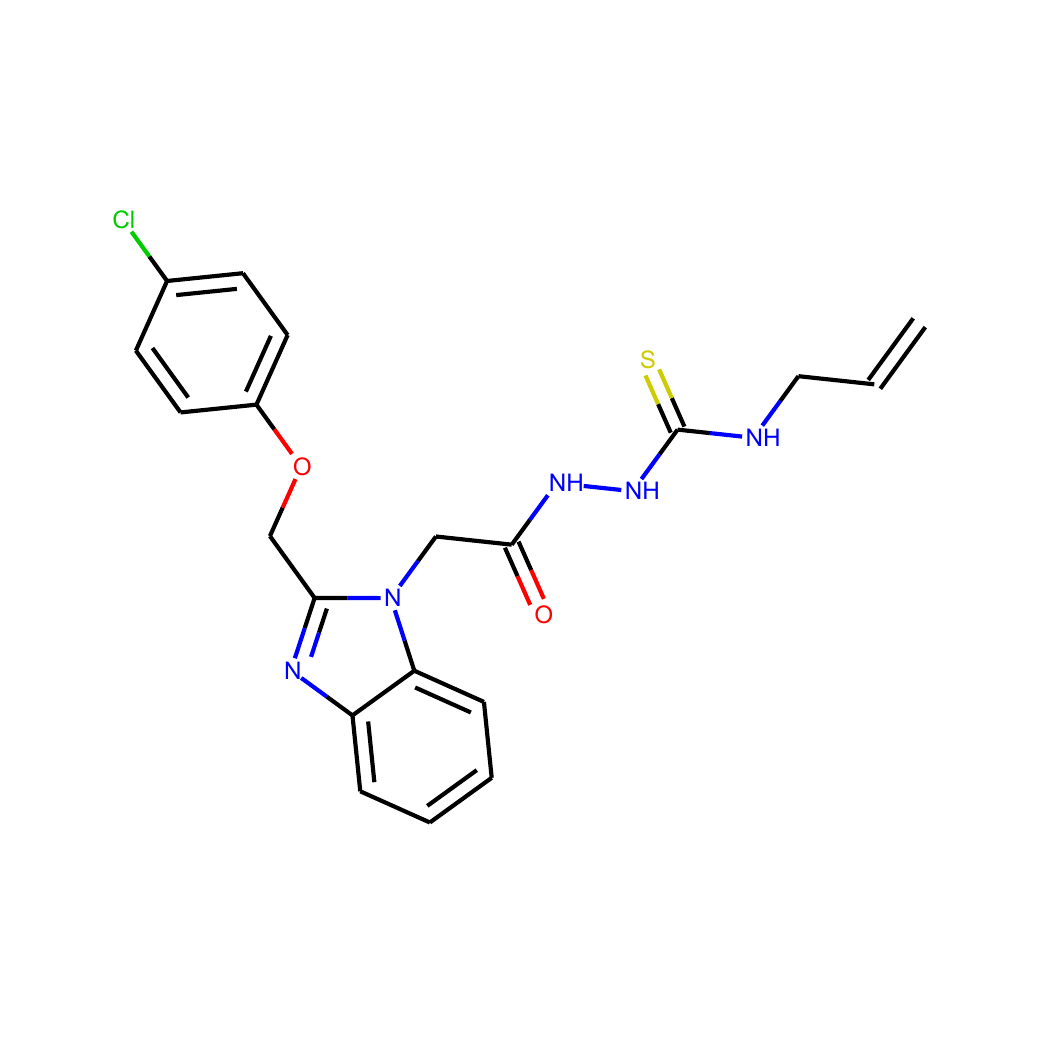} \\
\href{https://zinc.docking.org/substances/ZINC000806591744/}{ZINC000806591744} & \href{https://zinc.docking.org/substances/ZINC000178971373/}{ZINC000178971373} & \href{https://zinc.docking.org/substances/ZINC000000155607/}{ZINC000000155607} & \href{https://zinc.docking.org/substances/ZINC000016317677/}{ZINC000016317677} \\
\includegraphics[width=32mm]{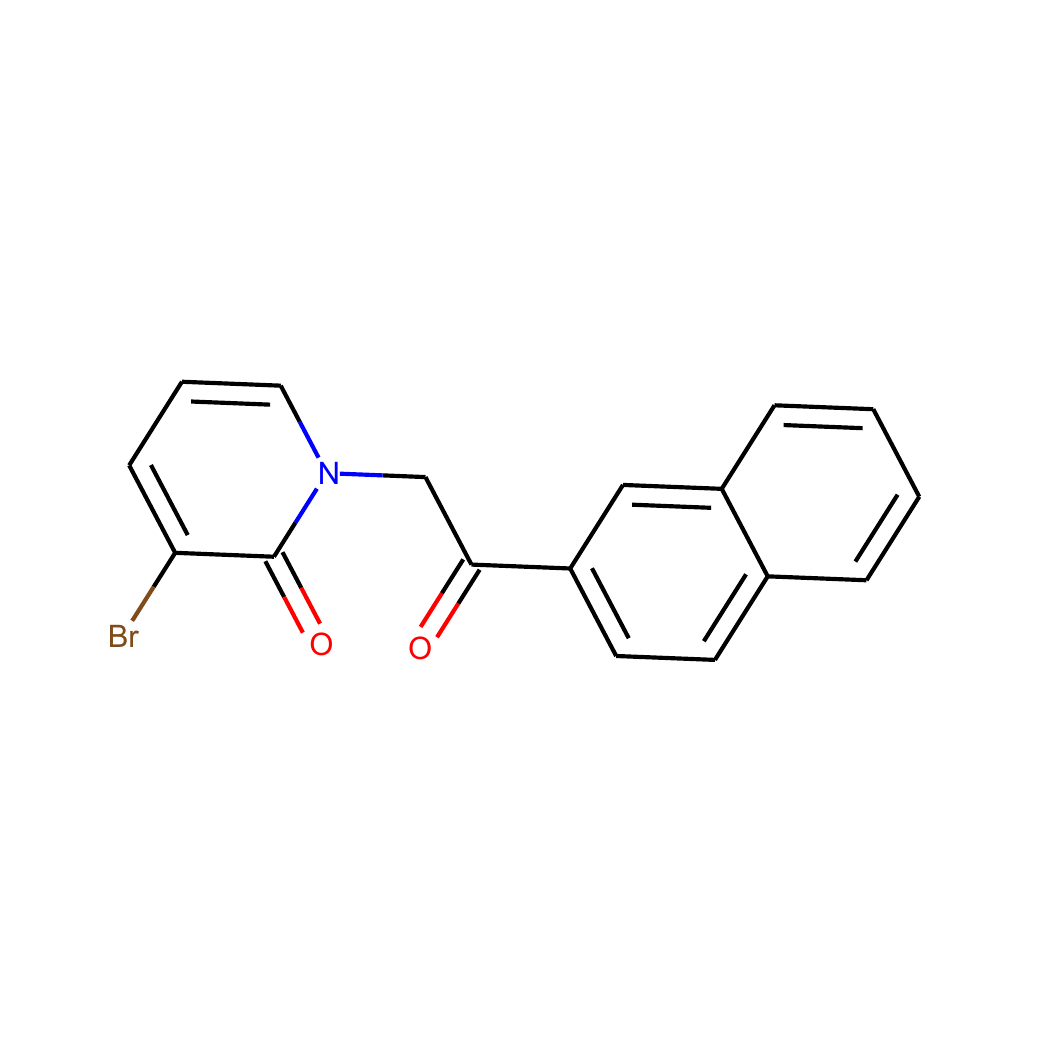} & \includegraphics[width=32mm]{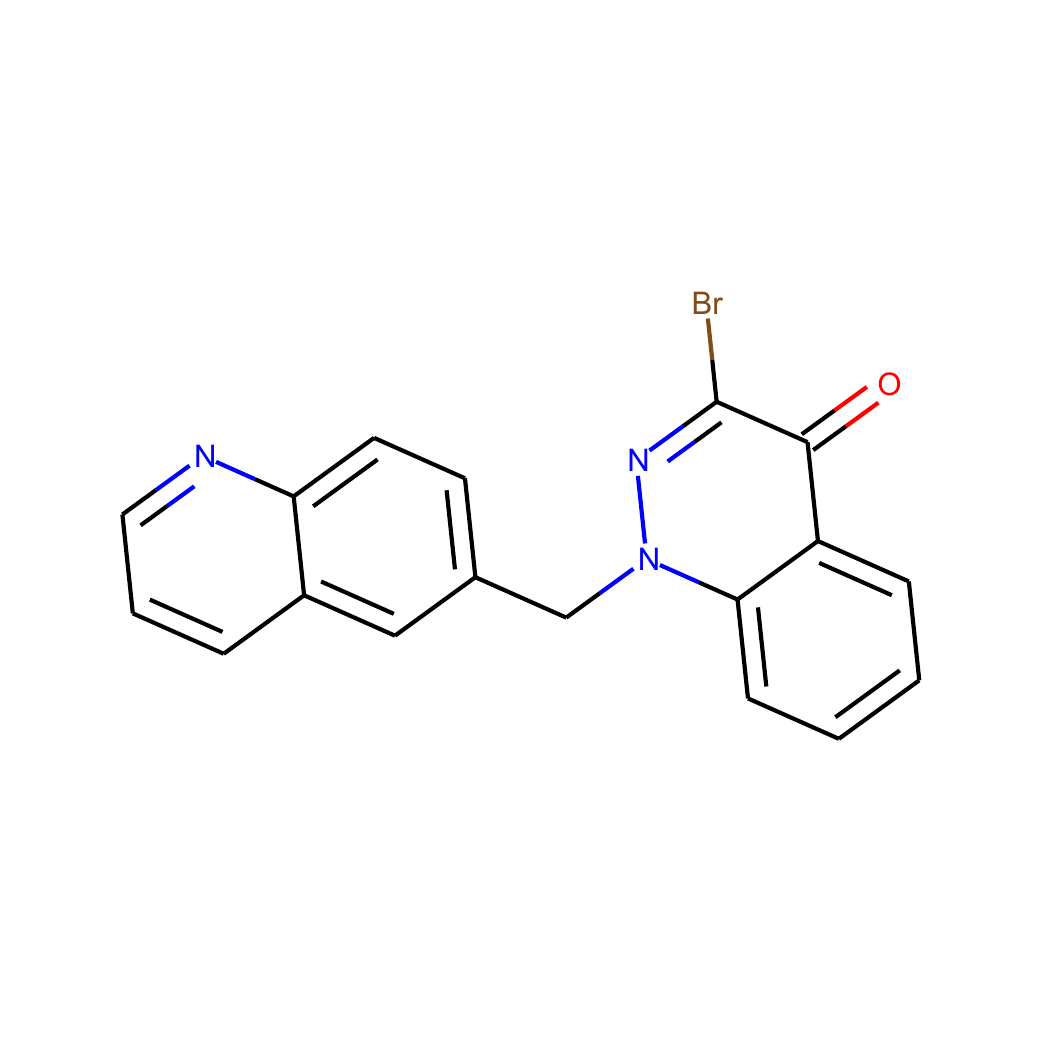} & \includegraphics[width=32mm]{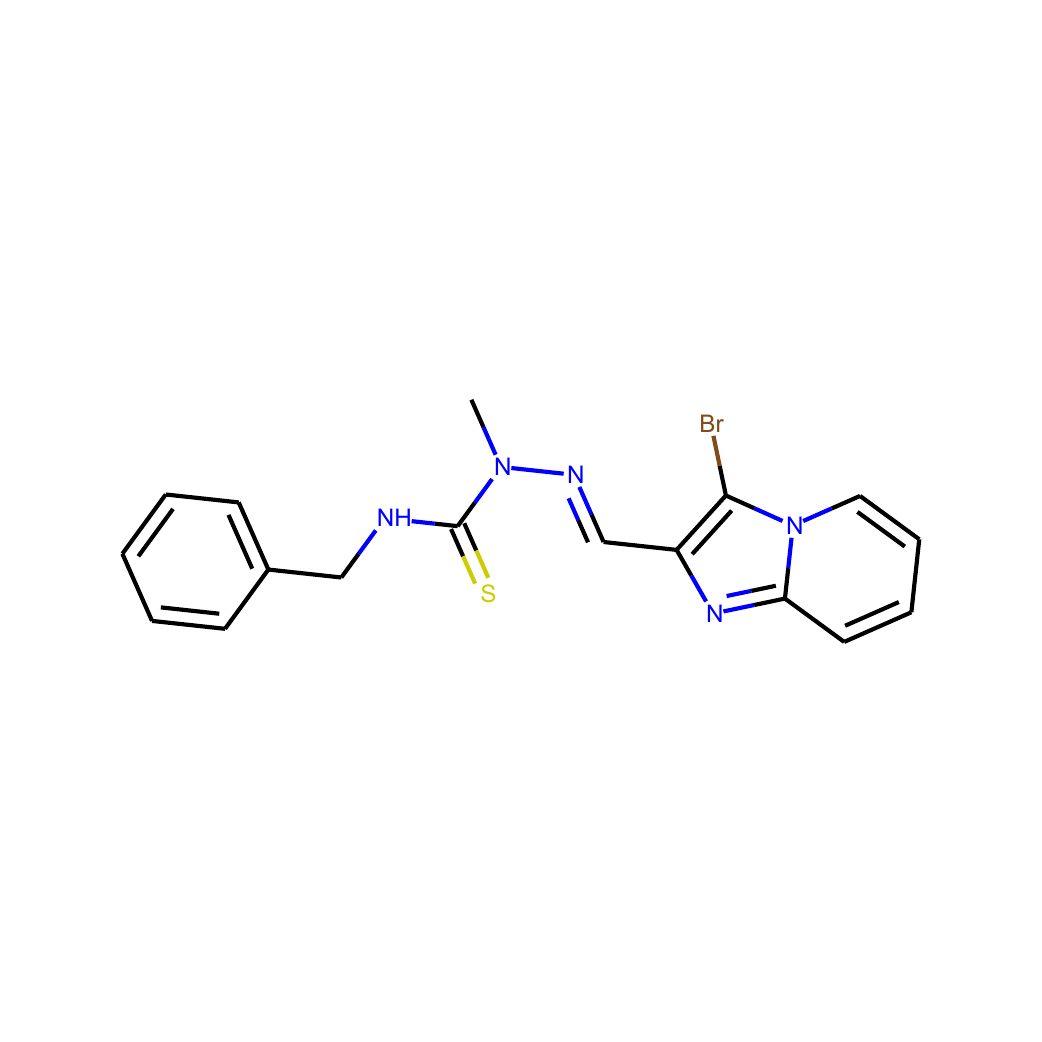} & \includegraphics[width=32mm]{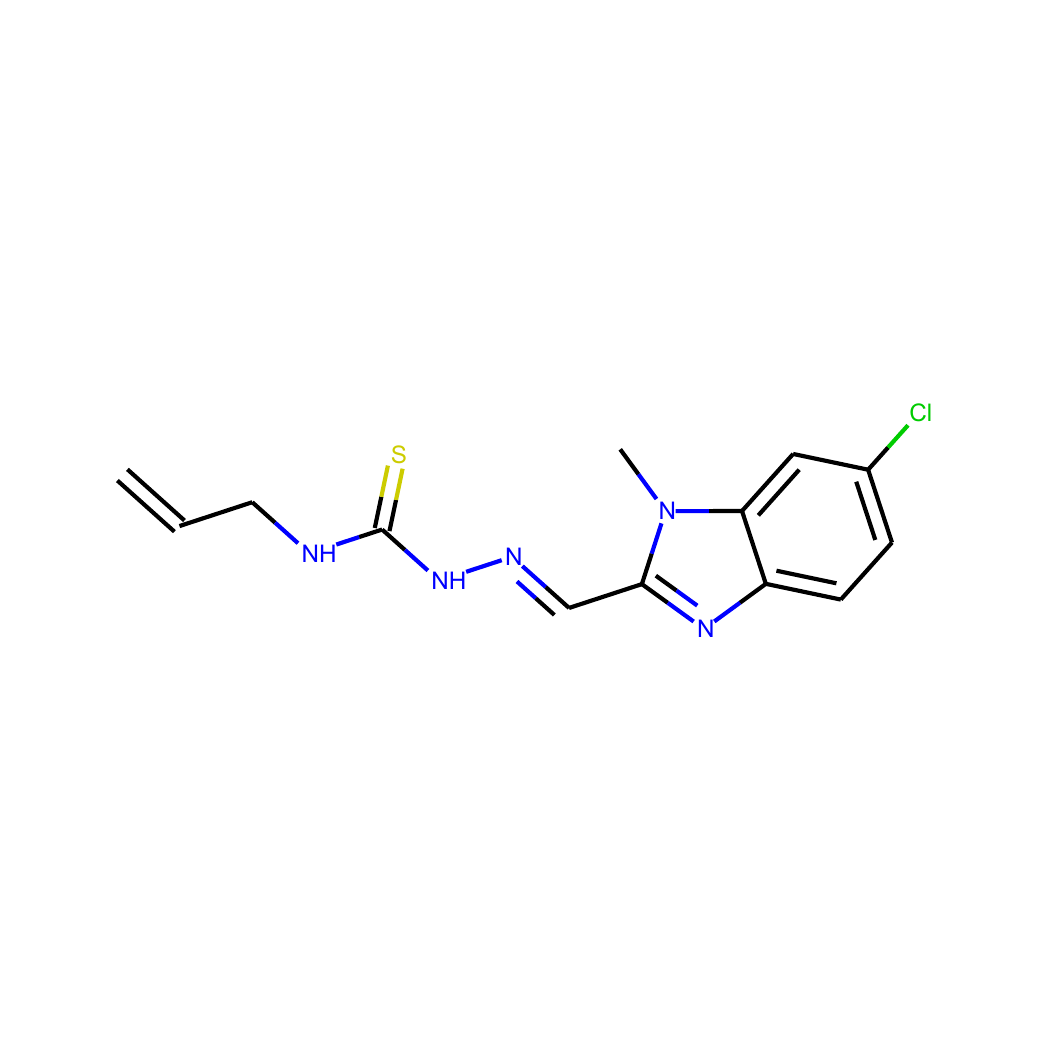} \\
\href{https://zinc.docking.org/substances/ZINC000193073749/}{ZINC000193073749} & \href{https://zinc.docking.org/substances/ZINC000769846795/}{ZINC000769846795} & \href{https://zinc.docking.org/substances/ZINC000755523869/}{ZINC000755523869} & \href{https://zinc.docking.org/substances/ZINC000763345954/}{ZINC000763345954} \\
\includegraphics[width=32mm]{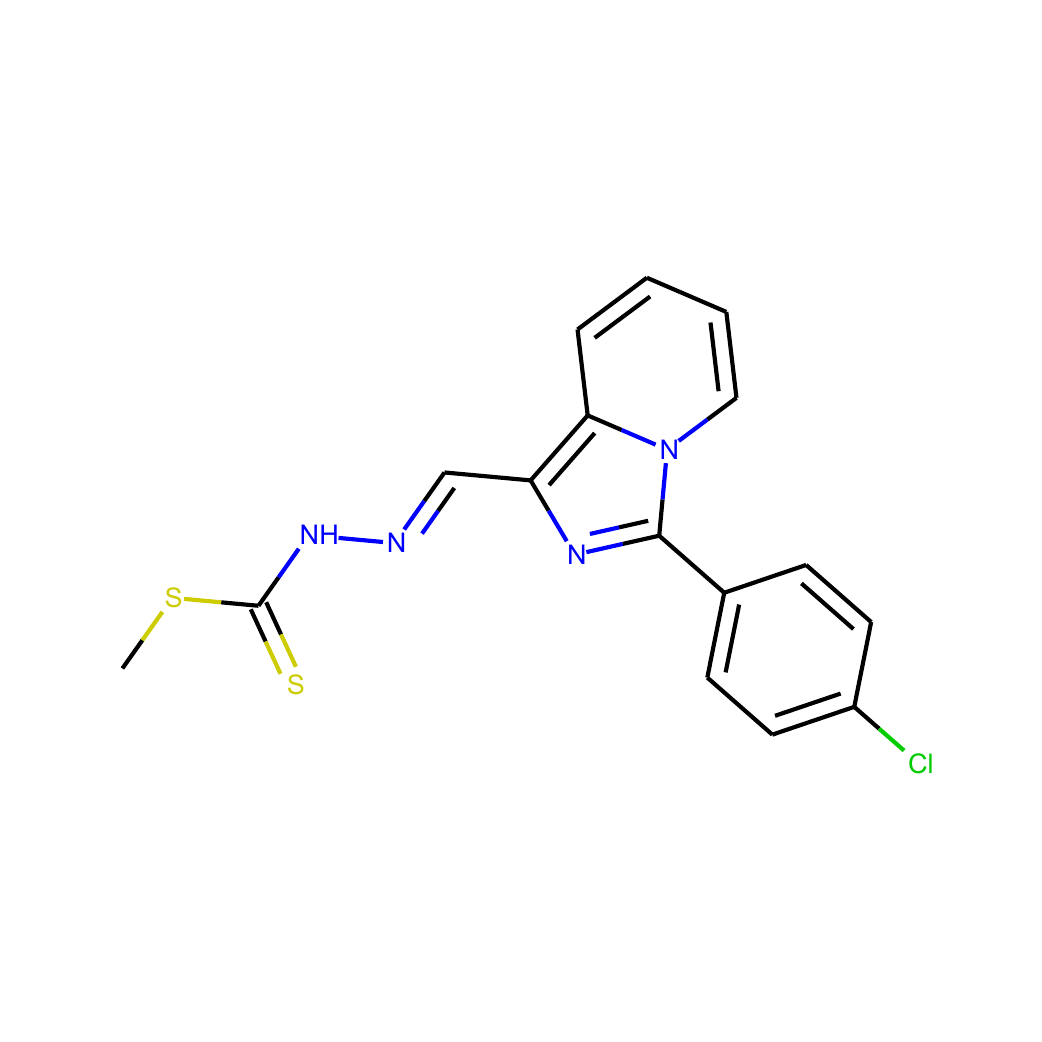} & \includegraphics[width=32mm]{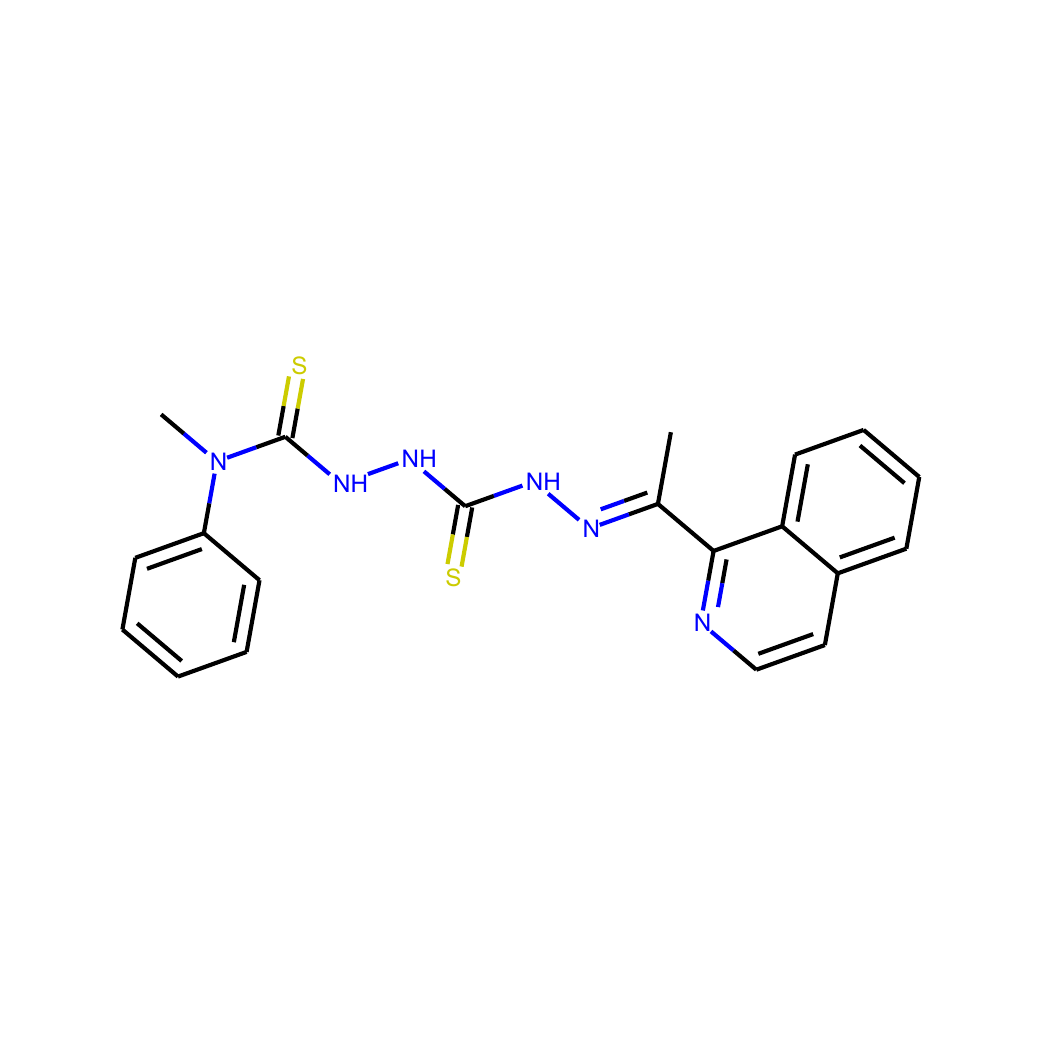} & \includegraphics[width=32mm]{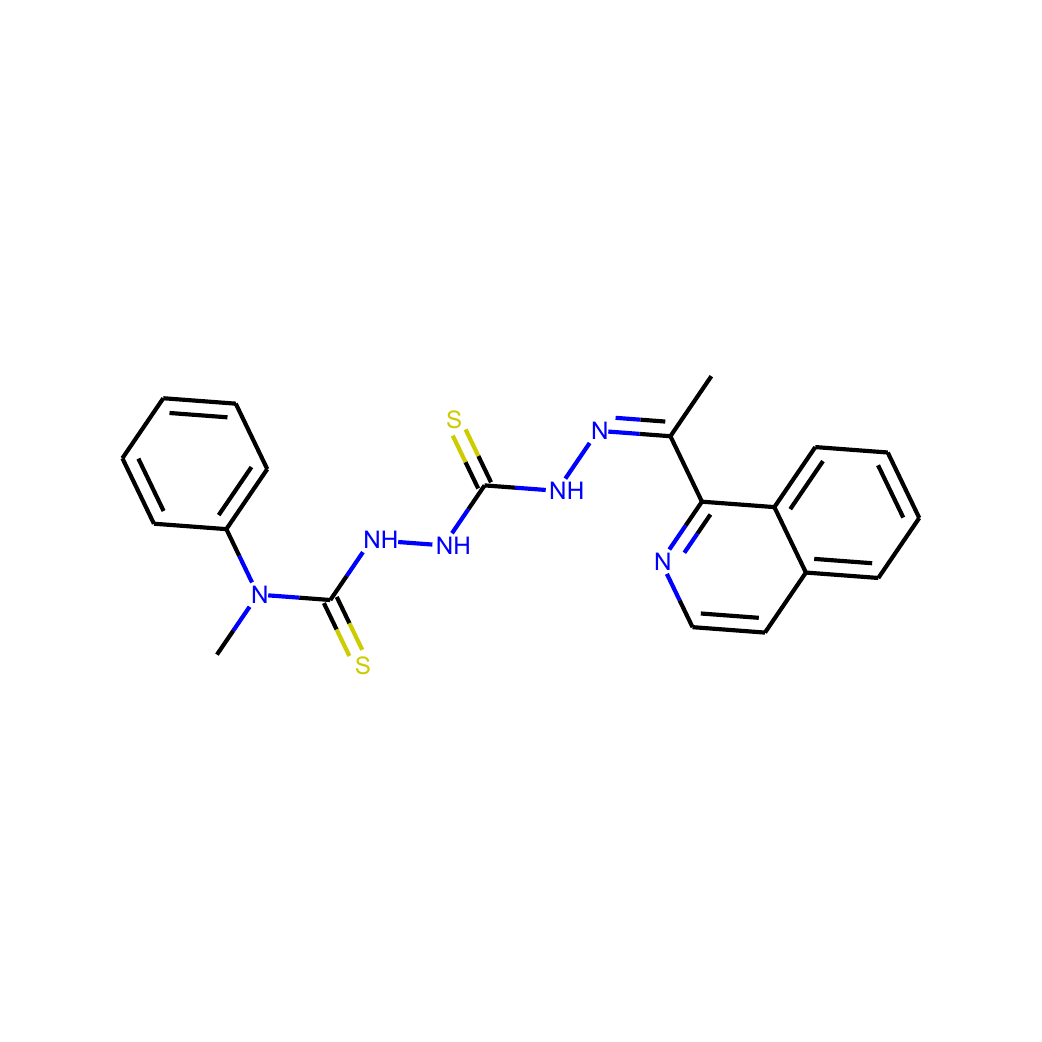} & \includegraphics[width=32mm]{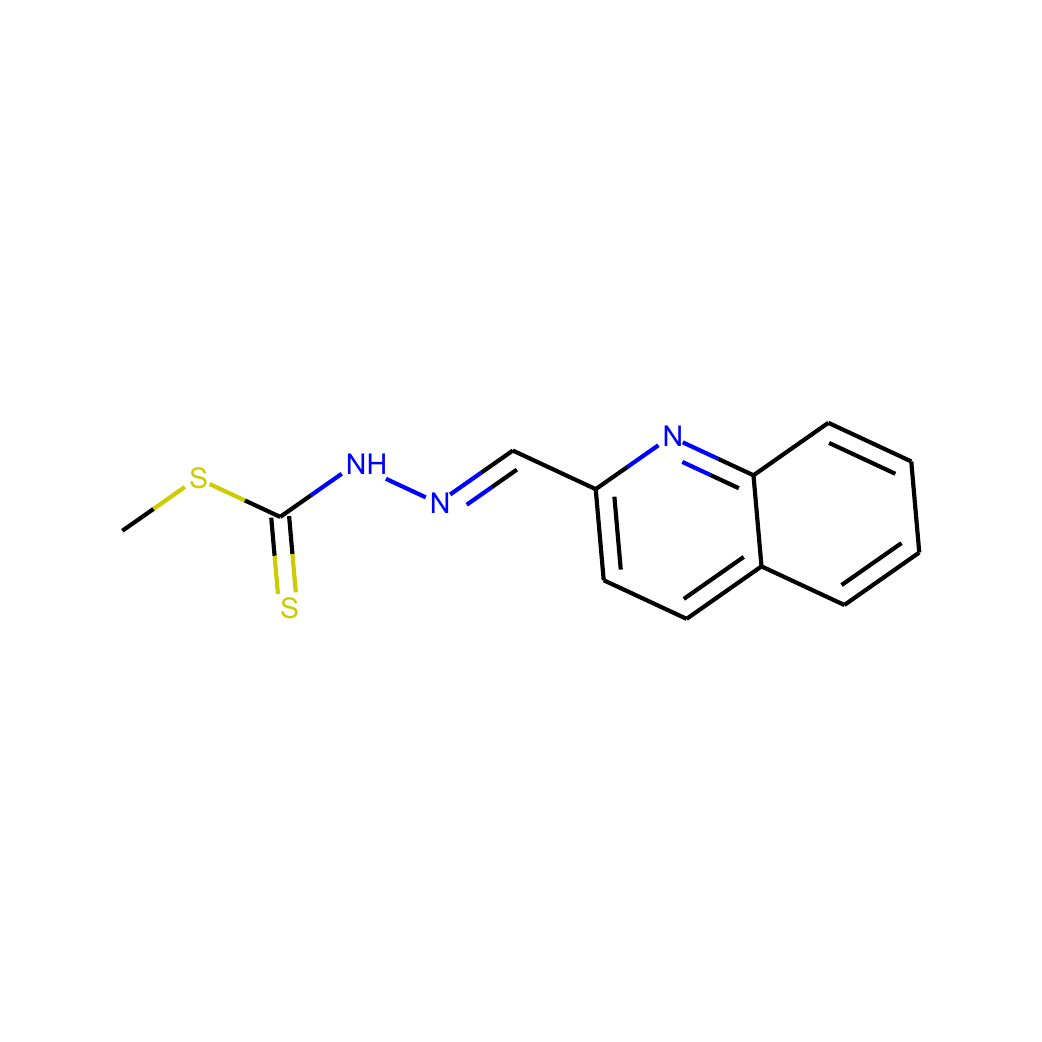} \\
\href{https://zinc.docking.org/substances/ZINC000001448699/}{ZINC000001448699} & \href{https://zinc.docking.org/substances/ZINC000016940508/}{ZINC000016940508} & \href{https://zinc.docking.org/substances/ZINC000005486767/}{ZINC000005486767} & \href{https://zinc.docking.org/substances/ZINC000005527649/}{ZINC000005527649} \\
\includegraphics[width=32mm]{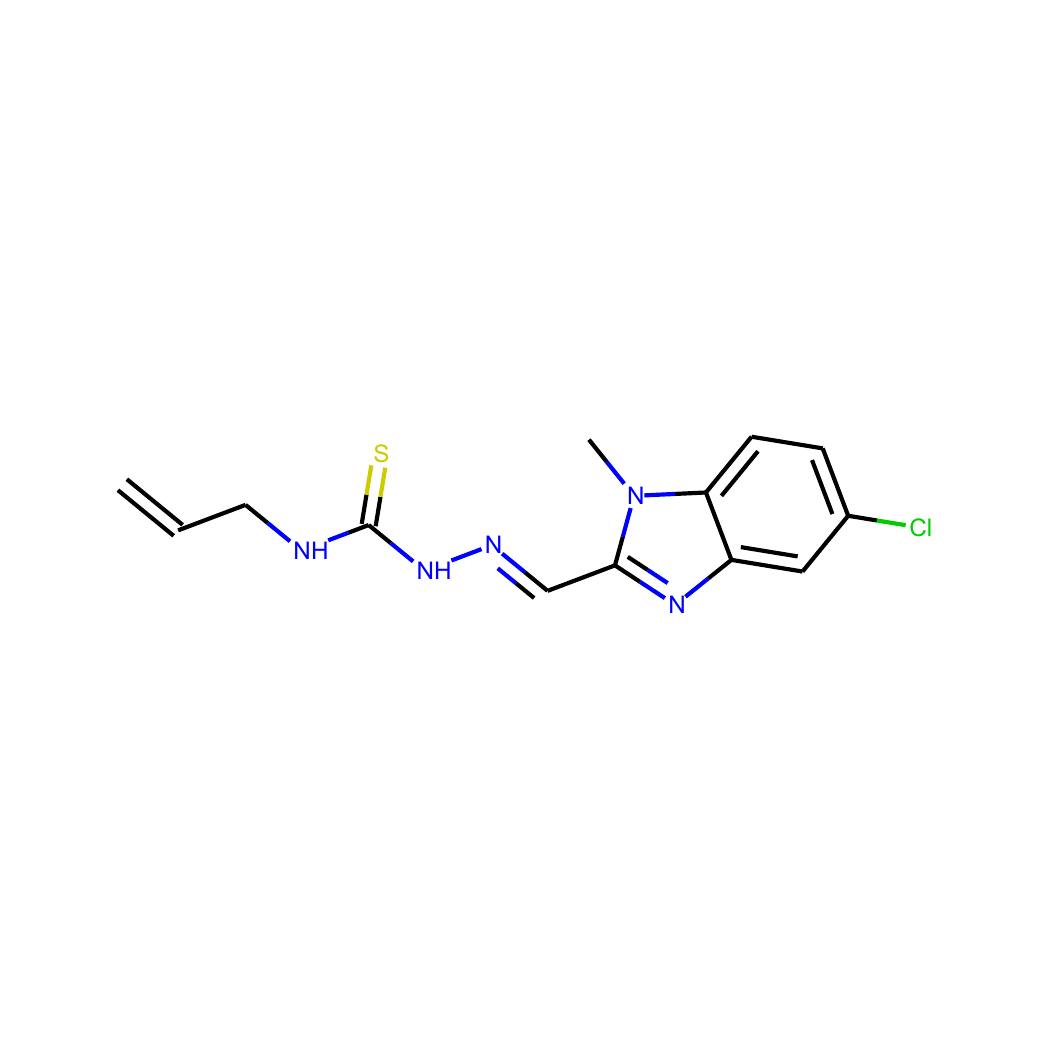} & \includegraphics[width=32mm]{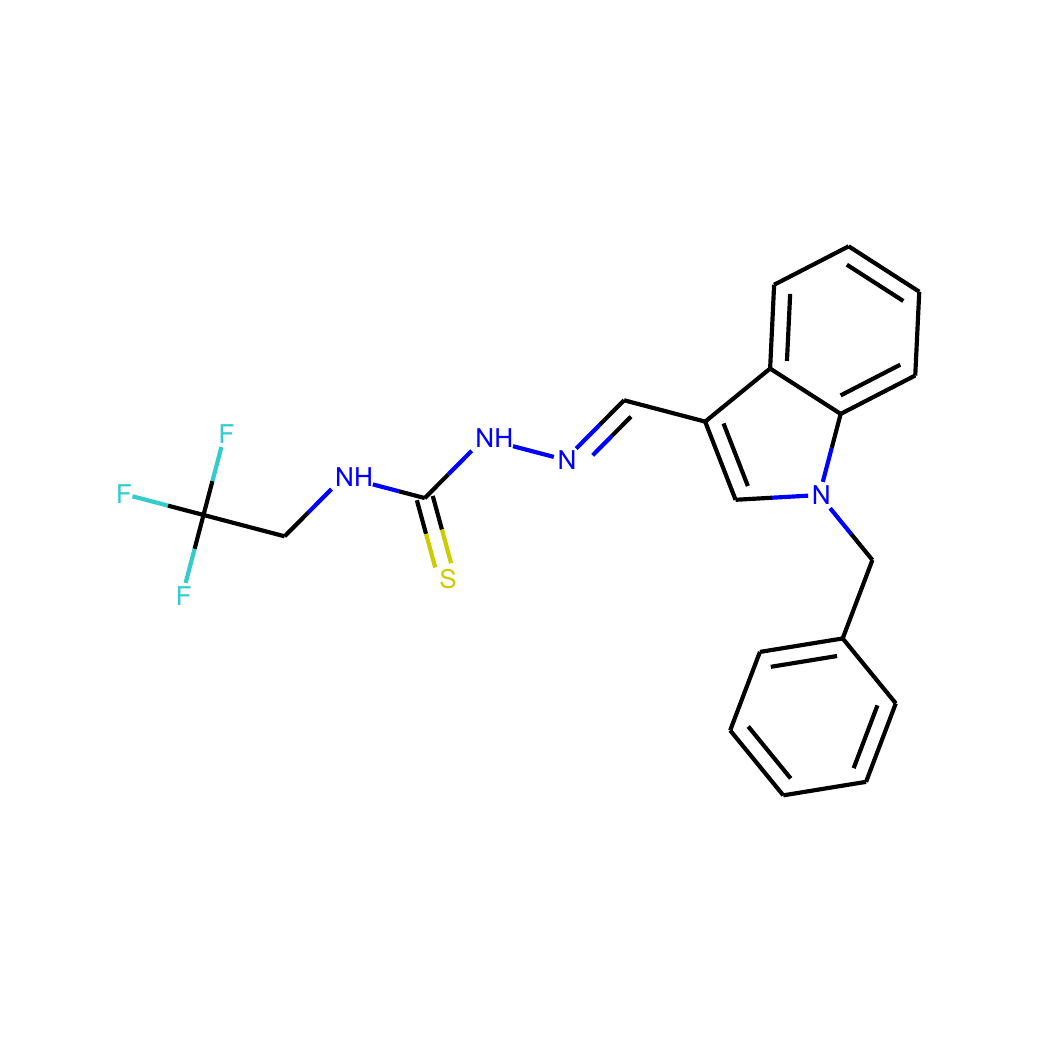} & \includegraphics[width=32mm]{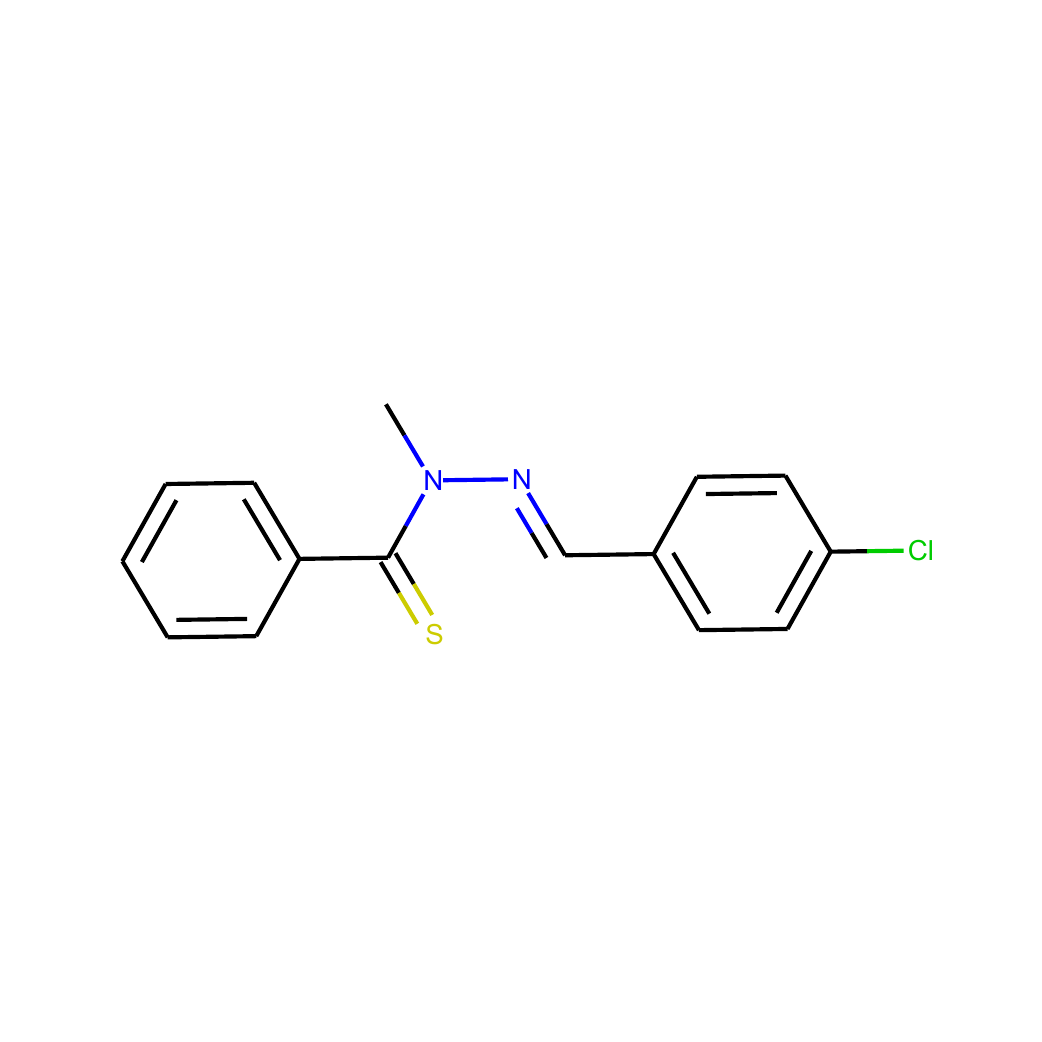} & \includegraphics[width=32mm]{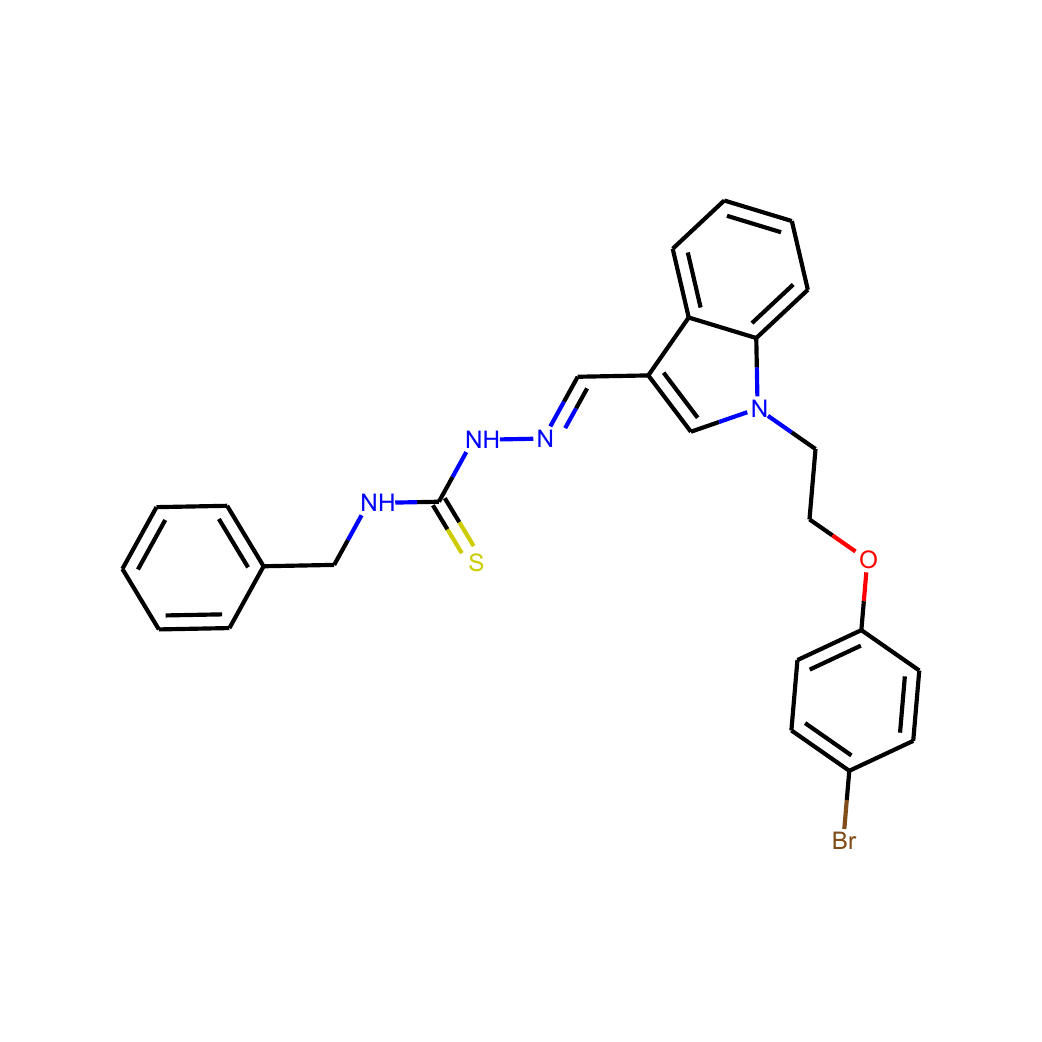} \\
\href{https://zinc.docking.org/substances/ZINC000755497029/}{ZINC000755497029} & \href{https://zinc.docking.org/substances/ZINC000746495682/}{ZINC000746495682} & \href{https://zinc.docking.org/substances/ZINC000005719506/}{ZINC000005719506} & \href{https://zinc.docking.org/substances/ZINC000002149503/}{ZINC000002149503} \\
    
    \end{tabular}
    \caption{Graphical representation of the top-ranked molecules by ChemAI. \label{fig:molecules}}
\end{figure*}

\paragraph{Consensus ranking.}
We developed a library of compounds which is enriched
for molecules with the ability to inhibit both proteases of the SARS-CoV-2. 
In order to score the multi-target effect, we calculated a consensus score for 
each molecule as the average rank 
of the predictions over the four selected
assays (see Table~\ref{tab:assays}). We then 
ranked all compounds by this consensus score. 
For each of the top-ranked compounds, we also calculated 
their minimal distance to
actives in the training set to be able to identify novel
chemical structures. Furthermore, for each compound we also 
report its number of potential toxic effects \citep{mayr2016deeptox}.

For the distance metric, we used the Jaccard distance based 
on binary ECFP4 fingerprints folded to a length of 1024, which 
yields values in the interval $[0,1]$. For potential toxic effects,
we used 75 output units of ChemAI with high predictive quality,
concretely an area under ROC-curve (AUC) larger than $0.80$, and 
counted how many of those output units indicated a toxic effect. This
value is reported in Table~\ref{tab:some_label} (column ``tox''). 
Furthermore, we report the clinical toxicity probability 
predicted by an independent
multitask neural network fitted on the 
ClinTox dataset \citep{wu_moleculenet_2018}. These probability
values are calibrated by Platt scaling \citep{platt1999probabilistic}
and reported in Table~\ref{tab:some_label} (column ``ct''). 
The additional information contained in these values can be used
to obtain a refined ranking for testing the molecules.

We implemented the overall process as a two-step
approach. In the first step, we reduced the ZINC
database of one billion molecules to a smaller set,
where we kept all molecules that exhibited some
predicted activity on any of the four assays
(precisely, at least one of the predictions
had to reside in the top-1\% quantile). 
In this way, we obtained an intermediate dataset of
5,672,501 molecules. For those molecules, the consensus
score, the toxicity flags and the distance to known 
actives were calculated. In the second step, we reduced
the dataset to the top-ranked 30,000 molecules by the
consensus score.

\paragraph{Results.} 
With the abovementioned approach, we assembled a
library of potential inhibitors of SARS-Cov-2. 
We report three metrics for each compound: 
a) predicted inhibitory effect of 
SARS-CoV proteases b) potential toxicities
and c) distance to known actives. This led to a ranked 
list of compounds of which we provide the top 30,000 as
a screening library. The top-ranked molecules are given
in Table~\ref{tab:some_label} and Figure~\ref{fig:molecules}.

We also checked whether molecules suggested by other publications
can be confirmed by ChemAI. Overall, some suggested molecules
show at least mild predicted activity against SARS-CoV (see 
Table~\ref{tab:overlapping}). 

\paragraph{Drug repurposing of DrugBank molecules.}
With the same procedure employed for ZINC, we also screened 
the DrugBank \citep{wishart2018drugbank}, a database of 
$\approx$10,000 drugs. Again, we predicted inhibitory effects 
on the two viral proteases using ChemAI, and then calculated
a consensus score for each drug. In this way, we obtained 
a ranked list of drugs that could be potential SARS-Cov 
inhibitors, which could be fast ways to therapies via 
drug repurposing \citep{ashburn2004drug}. We provide this 
list freely for research institutions (see~Availability).

\paragraph{Discussion.}
In this work, we presented the construction of 
a screening library of small molecules that are potential 
inhibitors of SARS-CoV-2. Our ligand-based approach uses
a neural network trained to predict the outcomes of bioassays.
From this multi-task models, four tasks have been selected to
predict the inhibitory potential against SARS-CoV-1. A consensus
between these predictions was used to rank compounds 
from the ZINC database, of which the 30,000 top-ranked are reported.

The approach is limited by the predictive quality of the underlying
machine learning method,
evaluated via AUC and leading to values in the 
range of $0.69$ to $0.78$. While these results are very promising, 
improved data quality, larger amount of data or machine-learning
approaches could lead to increased predictive performance and quality
of the library. A promising direction is also to enrich the 
representation of molecules
via already available biological 
modalities \citep{simm2018repurposing,hofmarcher2019accurate}.

We expect that the data for SARS-CoV-1 already has high predictive power 
for inhibitory effects of compounds on SARS-CoV-2. 
However, the current predictions can be further adjusted toward SARS-CoV-2 via 
transfer-learning and the incorporation of new data from SARS-CoV-2. 
In particular, few shot learning may be utilized for the first 
measurements for SARS-CoV-2 thus adjusting the multi-task model toward SARS-CoV-2.



\paragraph{Availability}
The library of molecules is 
available at \url{https://github.com/ml-jku/sars-cov-inhibitors-chemai}.

\section*{Acknowledgements}
Funding by the Institute for Machine Learning (JKU). 
All authors contributed equally to this work. 

\bibliography{references}

\begin{thebibliography}{37}
\providecommand{\natexlab}[1]{#1}
\providecommand{\url}[1]{\texttt{#1}}
\expandafter\ifx\csname urlstyle\endcsname\relax
  \providecommand{\doi}[1]{doi: #1}\else
  \providecommand{\doi}{doi: \begingroup \urlstyle{rm}\Url}\fi

\bibitem[Ashburn \& Thor(2004)Ashburn and Thor]{ashburn2004drug}
Ashburn, T.~T. and Thor, K.~B.
\newblock Drug repositioning: identifying and developing new uses for existing
  drugs.
\newblock \emph{Nature reviews Drug discovery}, 3\penalty0 (8):\penalty0
  673--683, 2004.

\bibitem[Chen et~al.(2020)Chen, Yiu, and Wong]{chen2020prediction}
Chen, Y.~W., Yiu, C.-P.~B., and Wong, K.-Y.
\newblock Prediction of the sars-cov-2 (2019-ncov) 3c-like protease (3cl pro)
  structure: virtual screening reveals velpatasvir, ledipasvir, and other drug
  repurposing candidates.
\newblock \emph{F1000Research}, 9, 2020.

\bibitem[Collison(2019)]{collison2019two}
Collison, J.
\newblock Two targets are better than one.
\newblock \emph{Nature Reviews Rheumatology}, 15\penalty0 (7):\penalty0
  386--386, 2019.

\bibitem[Fischer et~al.(2020)Fischer, Sellner, Neranjan, Lill, and
  Smie{\v{s}}ko]{fischer2020inhibitors}
Fischer, A., Sellner, M., Neranjan, S., Lill, M.~A., and Smie{\v{s}}ko, M.
\newblock Inhibitors for novel coronavirus protease identified by virtual
  screening of 687 million compounds.
\newblock 2020.

\bibitem[Gaulton et~al.(2017)Gaulton, Hersey, Nowotka, Bento, Chambers, Mendez,
  Mutowo, Atkinson, Bellis, Cibri{\'a}n-Uhalte, et~al.]{gaulton2017chembl}
Gaulton, A., Hersey, A., Nowotka, M., Bento, A.~P., Chambers, J., Mendez, D.,
  Mutowo, P., Atkinson, F., Bellis, L.~J., Cibri{\'a}n-Uhalte, E., et~al.
\newblock The chembl database in 2017.
\newblock \emph{Nucleic acids research}, 45\penalty0 (D1):\penalty0 D945--D954,
  2017.

\bibitem[Glantz-Gashai et~al.(2017)Glantz-Gashai, Meirson, Reuveni, and
  Samson]{gashai2017virtualscreening}
Glantz-Gashai, Y., Meirson, T., Reuveni, E., and Samson, A.~O.
\newblock {{V}irtual screening for potential inhibitors of {M}cl-1
  conformations sampled by normal modes, molecular dynamics, and nuclear
  magnetic resonance}.
\newblock \emph{Drug Des Devel Ther}, 11:\penalty0 1803--1813, 2017.

\bibitem[Gorgulla et~al.(2020)Gorgulla, Boeszoermenyi, Wang, Fischer, Coote,
  Das, Malets, Radchenko, Moroz, Scott, et~al.]{gorgulla2020open}
Gorgulla, C., Boeszoermenyi, A., Wang, Z.-F., Fischer, P.~D., Coote, P., Das,
  K. M.~P., Malets, Y.~S., Radchenko, D.~S., Moroz, Y.~S., Scott, D.~A., et~al.
\newblock An open-source drug discovery platform enables ultra-large virtual
  screens.
\newblock \emph{Nature}, pp.\  1--8, 2020.

\bibitem[Haider et~al.(2020)Haider, Subhani, Farooq, Ishaq, Khalid, Khan, and
  Niazi]{haider2020silico}
Haider, Z., Subhani, M.~M., Farooq, M.~A., Ishaq, M., Khalid, M., Khan, R.
  S.~A., and Niazi, A.~K.
\newblock In silico discovery of novel inhibitors against main protease (mpro)
  of sars-cov-2 using pharmacophore and molecular docking based virtual
  screening from zinc database.
\newblock 2020.

\bibitem[Hochreiter \& Schmidhuber(1997)Hochreiter and
  Schmidhuber]{hochreiter1997long}
Hochreiter, S. and Schmidhuber, J.
\newblock Long short-term memory.
\newblock \emph{Neural computation}, 9\penalty0 (8):\penalty0 1735--1780, 1997.

\bibitem[Hochreiter et~al.(2018)Hochreiter, Klambauer, and
  Rarey]{hochreiter2018machine}
Hochreiter, S., Klambauer, G., and Rarey, M.
\newblock Machine learning in drug discovery.
\newblock \emph{Journal of Chemical Information and Modeling}, 58\penalty0
  (9):\penalty0 1723--1724, 2018.
\newblock \doi{10.1021/acs.jcim.8b00478}.
\newblock URL \url{https://doi.org/10.1021/acs.jcim.8b00478}.
\newblock PMID: 30109927.

\bibitem[Hofmarcher et~al.(2019)Hofmarcher, Rumetshofer, Clevert, Hochreiter,
  and Klambauer]{hofmarcher2019accurate}
Hofmarcher, M., Rumetshofer, E., Clevert, D.-A., Hochreiter, S., and Klambauer,
  G.
\newblock Accurate prediction of biological assays with high-throughput
  microscopy images and convolutional networks.
\newblock \emph{Journal of chemical information and modeling}, 59\penalty0
  (3):\penalty0 1163--1171, 2019.

\bibitem[Huang et~al.(2020)Huang, Tang, Wu, Zhang, Wang, Wang, Song, Zhai,
  Zhao, Yang, et~al.]{huang2020virtual}
Huang, A., Tang, X., Wu, H., Zhang, J., Wang, W., Wang, Z., Song, L., Zhai,
  M.-a., Zhao, L., Yang, H., et~al.
\newblock Virtual screening and molecular dynamics on blockage of key drug
  targets as treatment for covid-19 caused by sars-cov-2.
\newblock 2020.

\bibitem[Jin et~al.(2020)Jin, Du, Xu, Deng, Liu, Zhao, Zhang, Li, Zhang, Duan,
  et~al.]{jin2020structure}
Jin, Z., Du, X., Xu, Y., Deng, Y., Liu, M., Zhao, Y., Zhang, B., Li, X., Zhang,
  L., Duan, Y., et~al.
\newblock Structure-based drug design, virtual screening and high-throughput
  screening rapidly identify antiviral leads targeting covid-19.
\newblock \emph{bioRxiv}, 2020.

\bibitem[Kim et~al.(2016)Kim, Thiessen, Bolton, Chen, Fu, Gindulyte, Han, He,
  He, Shoemaker, et~al.]{kim2016pubchem}
Kim, S., Thiessen, P.~A., Bolton, E.~E., Chen, J., Fu, G., Gindulyte, A., Han,
  L., He, J., He, S., Shoemaker, B.~A., et~al.
\newblock Pubchem substance and compound databases.
\newblock \emph{Nucleic acids research}, 44\penalty0 (D1):\penalty0
  D1202--D1213, 2016.

\bibitem[Klambauer et~al.(2019)Klambauer, Hochreiter, and
  Rarey]{klambauer2019machine}
Klambauer, G., Hochreiter, S., and Rarey, M.
\newblock Machine learning in drug discovery.
\newblock \emph{Journal of Chemical Information and Modeling}, 59\penalty0
  (3):\penalty0 945--946, 2019.
\newblock \doi{10.1021/acs.jcim.9b00136}.
\newblock URL \url{https://doi.org/10.1021/acs.jcim.9b00136}.

\bibitem[Landrum(2006)]{2006rdkit}
Landrum, G.
\newblock {{RDKit}}: {{Open}}-source cheminformatics, 2006.
\newblock URL \url{http://www.rdkit.org}.

\bibitem[Ledford(2009)]{ledford2009one}
Ledford, H.
\newblock One drug, two targets, 2009.

\bibitem[Lim et~al.(2016)Lim, Roy, and Song]{lim2016zika}
Lim, L., Roy, A., and Song, J.
\newblock Identification of a zika ns2b-ns3pro pocket susceptible to allosteric
  inhibition by small molecules including qucertin rich in edible plants.
\newblock \emph{bioRxiv}, 2016.
\newblock \doi{10.1101/078543}.
\newblock URL \url{https://www.biorxiv.org/content/early/2016/10/01/078543}.

\bibitem[Macchiagodena et~al.(2020)Macchiagodena, Pagliai, and
  Procacci]{macchiagodena2020inhibition}
Macchiagodena, M., Pagliai, M., and Procacci, P.
\newblock Inhibition of the main protease 3cl-pro of the coronavirus disease 19
  via structure-based ligand design and molecular modeling.
\newblock \emph{arXiv preprint arXiv:2002.09937}, 2020.

\bibitem[Mayr et~al.(2016)Mayr, Klambauer, Unterthiner, and
  Hochreiter]{mayr2016deeptox}
Mayr, A., Klambauer, G., Unterthiner, T., and Hochreiter, S.
\newblock Deeptox: toxicity prediction using deep learning.
\newblock \emph{Frontiers in Environmental Science}, 3:\penalty0 80, 2016.

\bibitem[Mayr et~al.(2018)Mayr, Klambauer, Unterthiner, Steijaert, Wegner,
  Ceulemans, Clevert, and Hochreiter]{mayr2018large}
Mayr, A., Klambauer, G., Unterthiner, T., Steijaert, M., Wegner, J.~K.,
  Ceulemans, H., Clevert, D.-A., and Hochreiter, S.
\newblock Large-scale comparison of machine learning methods for drug target
  prediction on chembl.
\newblock \emph{Chemical science}, 9\penalty0 (24):\penalty0 5441--5451, 2018.

\bibitem[Platt et~al.(1999)]{platt1999probabilistic}
Platt, J. et~al.
\newblock Probabilistic outputs for support vector machines and comparisons to
  regularized likelihood methods.
\newblock \emph{Advances in large margin classifiers}, 10\penalty0
  (3):\penalty0 61--74, 1999.

\bibitem[Preuer et~al.(2018)Preuer, Renz, Unterthiner, Hochreiter, and
  Klambauer]{preuer2018frechet}
Preuer, K., Renz, P., Unterthiner, T., Hochreiter, S., and Klambauer, G.
\newblock Fr{\'e}chet chemnet distance: a metric for generative models for
  molecules in drug discovery.
\newblock \emph{Journal of chemical information and modeling}, 58\penalty0
  (9):\penalty0 1736--1741, 2018.

\bibitem[Preuer et~al.(2019)Preuer, Klambauer, Rippmann, Hochreiter, and
  Unterthiner]{preuer2019interpretable}
Preuer, K., Klambauer, G., Rippmann, F., Hochreiter, S., and Unterthiner, T.
\newblock Interpretable deep learning in drug discovery.
\newblock In \emph{Explainable AI: Interpreting, Explaining and Visualizing
  Deep Learning}, pp.\  331--345. Springer, 2019.

\bibitem[Ruan et~al.(2020)Ruan, Liu, Guo, He, Huang, Jia, and
  Yang]{ruan2020potential}
Ruan, Z., Liu, C., Guo, Y., He, Z., Huang, X., Jia, X., and Yang, T.
\newblock Potential inhibitors targeting rna-dependent rna polymerase activity
  (nsp12) of sars-cov-2.
\newblock 2020.

\bibitem[Senathilake et~al.(2020)Senathilake, Samarakoon, and
  Tennekoon]{senathilake2020virtual}
Senathilake, K., Samarakoon, S., and Tennekoon, K.
\newblock Virtual screening of inhibitors against spike glycoprotein of 2019
  novel corona virus: a drug repurposing approach.
\newblock 2020.

\bibitem[Simm et~al.(2018)Simm, Klambauer, Arany, Steijaert, Wegner, Gustin,
  Chupakhin, Chong, Vialard, Buijnsters, et~al.]{simm2018repurposing}
Simm, J., Klambauer, G., Arany, A., Steijaert, M., Wegner, J.~K., Gustin, E.,
  Chupakhin, V., Chong, Y.~T., Vialard, J., Buijnsters, P., et~al.
\newblock Repurposing high-throughput image assays enables biological activity
  prediction for drug discovery.
\newblock \emph{Cell chemical biology}, 25\penalty0 (5):\penalty0 611--618,
  2018.

\bibitem[Sterling \& Irwin(2015)Sterling and Irwin]{sterling2015zinc}
Sterling, T. and Irwin, J.~J.
\newblock Zinc 15--ligand discovery for everyone.
\newblock \emph{Journal of chemical information and modeling}, 55\penalty0
  (11):\penalty0 2324--2337, 2015.

\bibitem[Ton et~al.(2020)Ton, Gentile, Hsing, Ban, and Cherkasov]{ton2020rapid}
Ton, A.-T., Gentile, F., Hsing, M., Ban, F., and Cherkasov, A.
\newblock Rapid identification of potential inhibitors of sars-cov-2 main
  protease by deep docking of 1.3 billion compounds.
\newblock \emph{Molecular Informatics}, 2020.

\bibitem[Wang et~al.(2020)Wang, Zhao, Chen, and Hong]{wang2020virtual}
Wang, Q., Zhao, Y., Chen, X., and Hong, A.
\newblock Virtual screening of approved clinic drugs with main protease
  (3clpro) reveals potential inhibitory effects on sars-cov-2.
\newblock 2020.

\bibitem[Weininger(1988)]{weininger1988smiles}
Weininger, D.
\newblock {SMILES}, a chemical language and information system. 1. introduction
  to methodology and encoding rules.
\newblock \emph{Journal of chemical information and computer sciences},
  28\penalty0 (1):\penalty0 31--36, 1988.

\bibitem[Wishart et~al.(2018)Wishart, Feunang, Guo, Lo, Marcu, Grant, Sajed,
  Johnson, Li, Sayeeda, et~al.]{wishart2018drugbank}
Wishart, D.~S., Feunang, Y.~D., Guo, A.~C., Lo, E.~J., Marcu, A., Grant, J.~R.,
  Sajed, T., Johnson, D., Li, C., Sayeeda, Z., et~al.
\newblock Drugbank 5.0: a major update to the drugbank database for 2018.
\newblock \emph{Nucleic acids research}, 46\penalty0 (D1):\penalty0
  D1074--D1082, 2018.

\bibitem[Wu et~al.(2020)Wu, Liu, Yang, Zhang, Zhong, Wang, Wang, Xu, Li, Li,
  Zheng, Chen, and Li]{wu2020analysis}
Wu, C., Liu, Y., Yang, Y., Zhang, P., Zhong, W., Wang, Y., Wang, Q., Xu, Y.,
  Li, M., Li, X., Zheng, M., Chen, L., and Li, H.
\newblock Analysis of therapeutic targets for sars-cov-2 and discovery of
  potential drugs by computational methods.
\newblock \emph{Acta Pharmaceutica Sinica B}, 2020.
\newblock ISSN 2211-3835.
\newblock \doi{https://doi.org/10.1016/j.apsb.2020.02.008}.
\newblock URL
  \url{http://www.sciencedirect.com/science/article/pii/S2211383520302999}.

\bibitem[Wu et~al.(2018)Wu, Ramsundar, Feinberg, Gomes, Geniesse, Pappu,
  Leswing, and Pande]{wu_moleculenet_2018}
Wu, Z., Ramsundar, B., Feinberg, E., Gomes, J., Geniesse, C., Pappu, A.~S.,
  Leswing, K., and Pande, V.
\newblock {MoleculeNet}: {A} benchmark for molecular machine learning.
\newblock \emph{Chemical Science}, 9\penalty0 (2):\penalty0 513--530, 2018.
\newblock ISSN 2041-6520, 2041-6539.
\newblock \doi{10.1039/C7SC02664A}.
\newblock URL \url{http://xlink.rsc.org/?DOI=C7SC02664A}.

\bibitem[Zhang et~al.(2020)Zhang, Shen, Yan, Yan, and
  Cheng]{zhang2020discovery}
Zhang, J.-J., Shen, X., Yan, Y.-M., Yan, W., and Cheng, Y.-X.
\newblock Discovery of anti-sars-cov-2 agents from commercially available
  flavor via docking screening.
\newblock 2020.

\bibitem[Zhou et~al.(2020)Zhou, Hou, Shen, Huang, Martin, and
  Cheng]{zhou2020network}
Zhou, Y., Hou, Y., Shen, J., Huang, Y., Martin, W., and Cheng, F.
\newblock Network-based drug repurposing for novel coronavirus
  2019-ncov/sars-cov-2.
\newblock \emph{Cell Discovery}, 6\penalty0 (1):\penalty0 1--18, 2020.

\bibitem[Zhu et~al.(2020)Zhu, Wang, Yang, Zhang, Mu, Shi, Peng, Xu,
  et~al.]{zhu2020d3similarity}
Zhu, Z., Wang, X., Yang, Y., Zhang, X., Mu, K., Shi, Y., Peng, C., Xu, Z.,
  et~al.
\newblock D3similarity: A ligand-based approach for predicting drug targets and
  for virtual screening of active compounds against covid-19.
\newblock 2020.

\end{thebibliography}
\bibliographystyle{icml2020}

\appendix

\begin{table*}[]
    \centering
    \begin{tabular}{llp{6cm}l}
        \toprule
        ZINC ID & Trivial name(s) & Canonical SMILES &  Publications \\
        \midrule
        ZINC00057060 &           Melatonin &               COc1ccc2[nH]cc(CCNC(C)=O)c2c1 &                                    \citet{zhou2020network} \\
        ZINC03869685 &  Meletin, Quercetin &         O=c1c(O)c(-c2ccc(O)c(O)c2)oc2cc(O)cc(O)c12 &                                        \citet{lim2016zika} \\
        ZINC85537142 &         Aclarubicin & CC[C@@]1(O)C[C@H](O[C@H]2C [C@H](N(C)C)[C@H](O[C@H]3C[C@H] (O)[C@H](O[C@H]4CCC(=O)[C@H] (C)O4)[C@H](C)O3)[C@H](C)O2) c2c(cc3c(c2O)C(=O)c2c(O) cccc2C3=O)[C@H]1C(=O)OC &                             \citet{senathilake2020virtual} \\
        ZINC03794794 &        Mitoxantrone &  C1=CC(=C2C(=C1NCCNCCO) C(=O)C3=C(C=CC (=C3C2=O)O)O)NCCNCCO &                                    \citet{wang2020virtual} \\
        ZINC01668172 &                   - &        O=C(C[n+]1ccc2ccccc2c1) c1ccc2ccc3ccccc3c2c1 &  \citet{gashai2017virtualscreening} \\
        ZINC03830332 &                E155 &  C1=CC=C2C(=C1)C(=CC=C2S(=O) (=O)O)NN=C3C=C(C(=O)C (=NNC4=CC=C(C5=CC=CC=C54) S(=O)(=O)O)C3=O)CO &                             \citet{senathilake2020virtual} \\
        ZINC14879972 & Gar-936 & CN(C)c1cc(NC(=O)CNC(C) (C)C)c(O)c2c1C[C@H]1C [C@H]3[C@H](N(C)C)C(O) =C(C(N)=O)C(=O)[C@@] 3(O)C(O)=C1C2=O & \citet{wu2020analysis} \\
        ZINC00001645 & Magnolol & C=CCc1ccc(O)c(-c2cc (CC=C)ccc2O)c1 & \citet{wu2020analysis} \\
        ZINC00014036 & Piceatannol & Oc1cc(O)cc(/C=C/c2ccc(O)c(O)c2)c1 & \citet{wu2020analysis} \\
        ZINC16052277 & Doxycycline & C[C@H]1c2cccc(O)c2C(=O) C2=C(O)[C@]3(O)C(=O)C (C(=N)O)=C(O)[C@@H](N(C)C) [C@@H]3[C@@H](O)[C@@H]21 & \citet{wu2020analysis} \\
        ZINC3920266	& Idarubicin & CC(=O)[C@]1(O)Cc2c(O)c3 c(c(O)c2[C@@H](O[C@H]2C [C@H](N)[C@H](O)[C@H] (C)O2)C1)C(=O)c1ccccc1C3=O & \citet{wu2020analysis} \\

        \bottomrule
    \end{tabular}
    \caption{Compounds suggested in related publications
    for potential activity against SARS-CoV-2 and which also exhibit at least mild predicted activity against SARS-CoV proteases by ChemAI. 
    }
    \label{tab:overlapping}
\end{table*}

\end{document}